\documentclass[aps,preprint,showpacs,showkeys,preprintnumbers,amsmath,amssymb]{revtex4}

\usepackage{txfonts}
\usepackage{dcolumn}
\usepackage{mathrsfs}
\usepackage{bm}
\usepackage{amsmath,amssymb,epsfig,float}

\begin{document}

\title{Multipartite entanglement for open system in noninertial frames }

\author{Wenpin Zhang and Jiliang Jing\footnote{Corresponding author, Email: jljing@hunnu.edu.cn}}
\affiliation{ Department of Physics, and Key Laboratory of Low
Dimensional Quantum Structures \\ and Quantum Control of Ministry of
Education, Hunan Normal University, \\ Changsha, Hunan 410081,
People's Republic of China}

\vspace*{0.2cm}
\begin{abstract}
\vspace*{0.2cm}

Based on Greenberger-Horne-Zeilinger ($GHZ$) and $W$ initial states,
the tripartite entanglement of a fermionic system under the
amplitude damping channel and in depolarizing noise when two
subsystems accelerated is investigated. Unlike the case of two-qubit
system in which sudden death occurs easily, we find here that the
sudden death never occurs even all subsystems are under the noise
environment. We note that both acceleration and environment can
destroy the symmetry between the subsystems, but the effect of
environment is much stronger than that of acceleration. We also show
that an entanglement rebound process will take place when $P>0.75$
in the depolarizing noise and the larger acceleration will result in
the weaker rebound process.

\end{abstract}
\vspace*{1.cm} \pacs{03. 65. Ud,  03. 67. Mn,  04. 70.Dy}


\maketitle

\section{introduction}
Quantum entanglement plays as an important resource in quantum
computation \cite{1} , teleportation \cite{2}, dense coding \cite{3}
and cryptography \cite{4,5}. Since the environment is unavoidable in
practice, there were many meaningful works on the dynamics of
entanglement in two qubits which interacted with different kinds of
environments, and some important features of the entanglement such
as the entanglement sudden death \cite{6,7,8,9} and birth \cite{10}
were found. On the other hand, the relativistic quantum information
has been a focus of research over recent years for both conceptual
and experimental reasons. The studies of entanglement in noninertial
frames have shown that the Unruh or Hawking effect will influence
the degree of entanglement \cite{11,12,13,14,15} dramatically.
However, most of these works focused on the study of quantum
information in bipartite systems and only one of the subsystems
accelerated.

Along this line, in real quantum information tasks we have to
consider multipartite entanglement which interacts with different
kind of environments in inertial or noninertial frame. Recently, the
tripartite entanglement of scalar and Dirac fields in noninertial
frames were studied by Mi-Ra Hwang $et$ $al$. \cite{16} and Jieci
Wang $et$ $al$. \cite{17}. They showed that the tripartite
entanglement decreases with the increase of the acceleration, and all
the two-tangles equal to zero when one or two observers accelerated
for $GHZ$ initial state which is exactly the same as the two-tangles
obtained in the inertial frame.

In this paper we will discuss the  tripartite entanglement of Dirac
fields under the environment for amplitude damping channel and
depolarizing noise when two observers accelerated for $GHZ$ and $W$
initial states. Our setting consists of three observers: Alice, Bob
and Charlie. We assume that Alice is in an inertial frame, while Bob
and Charlie are in accelerated frames with the same uniformly
acceleration. We will focus our attention on how the environment and
acceleration influence the degree of tripartite entanglement. We
first let the inertial observer Alice under the environment, then
the noninertial observer Charlie under the environment, and at last
all of them under the environment. The Dirac fields, as shown in
Refs. \cite{18,19,20}, from an inertial perspective, can be
described as a superposition of Minkowski monochromatic modes
$|0\rangle_{M}=\bigotimes_i|0_{\omega_i}\rangle_{M} $ and
$|1\rangle_{M}=\bigotimes_i|1_{\omega_i}\rangle_{M} ~\forall i$,
with
\begin{eqnarray}\label{Eq.1}
|0_{\omega_{i}}\rangle_{M}&=& \cos r _{i}|0_{\omega_{i}}
\rangle_{I}|0_{\omega_{i}}\rangle _{II}+\sin
r_{i}|1_{\omega_{i}}\rangle_{I}|1_{\omega_{i}}\rangle _{II},\nonumber \\
|1_{\omega_{i}}\rangle_{M}&=& |1_{\omega_{i}}\rangle_{I}
|0_{\omega_{i}}\rangle _{II},
\end{eqnarray}
where $\cos r_{i}=(e^{-2\pi\omega_{i} c/a_{i}}+1)^{-1/2}$,  $a_{i}$
is the acceleration of the observer $i$ and $c$ is the acceleration
of the light. On account of the accelerated observer in Rindler
region $I$ are causally disconnected from region $II$, by tracing
over the inaccessible modes we will obtain a tripartite state and
then we can calculate tripartite entanglement of the 3-qubit states.

This paper is structured as follows. In Sec. II we will study the
environment and some measurements of tripartite entanglement. In
Sec. III we will discuss the tripartite entanglement of Dirac fields
when two observes are accelerated for $GHZ$ state and compare the
case of different subsystems under the environment. In Sec. IV we
will analyze the tripartite entanglement for the $W$ state under the
environment. Our work will be summarized in the last section.

\section{The ENVIRONMENT AND MEASURES}
Here we consider the local channel, in which all the subsystems
interact independently with its own environment and no communication
appears. If the local environment acts independently on subsystem's
state, the total evolution of these qubit systems can be expressed
as \cite{21}
\begin{eqnarray} \label{Eq.3}
L(\rho_{s})=\sum_{\mu ...\nu} M^{1}_\mu \otimes\cdots  \otimes
M^{N}_\nu \rho_{s} M_\mu^{1\dag}\otimes\cdots \otimes M_\nu^{N\dag},
\label{EvolKraus}
 \end{eqnarray}
where $M_{\mu}^{i}$ are the Kraus operators and $N$ is  the number
of the subsystems.

For the amplitude damping environment, we can take
\begin{eqnarray}
M_0^{i}=\left(\begin{array}{cc}
           1&0\\
           0&\sqrt{1-P_{i}}
           \end{array}\right),  &\; & M^{i}_1= \left(\begin{array}{cc}
                                           0&\sqrt{P_{i}}\\
                                           0&0
                                          \end{array}\right),
\label{Eq.4}
\end{eqnarray}
where $i=(1,2\cdots N)$, $\mu=(0,1)$.
This channel represents the dissipative interaction between
the qubit and its environment. The emblematic example
is given by the spontaneous emission of a photon by a
two-level atom into a zero-temperature environment of
electromagnetic-field modes.
A simple way to gain insight about this process is through
the corresponding quantum map \cite{21}
\begin{eqnarray}
\label{AmplitudeDampingMap}
|0\rangle_{S}|0\rangle_E&\rightarrow&
|0\rangle_{S}|0\rangle_E  \label{Eq.5}\; , \\
|1\rangle_{S}|0\rangle_E&\rightarrow&
\sqrt{1-P}|1\rangle_{S}|0\rangle_E +
\sqrt{P}|0\rangle_{S}|1\rangle_E  \label{Eq.6}\; .
\end{eqnarray}
Eq. (\ref{Eq.5}) shows that, if the system stays at $|0\rangle_{S}$,
both it and it's environment will not change at all.  Eq.
(\ref{Eq.6}) indicates that, if the system stays at $|1\rangle_{S}$,
the decay will exist in the system with probability $P$, and it can
also remain there with probability $(1-P)$.

For the depolarizing noise, due to the state is not stable
absolutely in this channel, the qubits will  have three mistakenly
flip in random. Assume that the three mistakenly flip take the same
probability then the responding quantum map becomes
\begin{eqnarray}\label{Eq.7}
|\Psi\rangle_{S}|0\rangle_E&\rightarrow&
\sqrt{1-P_{i}}|\Psi\rangle_{S}|0\rangle_E+
\sqrt{\frac{P_{i}}{3}}|\Psi\rangle_{S}|1\rangle_E+
\sqrt{\frac{P_{i}}{3}}|\Psi\rangle_{S}|2\rangle_E+
\sqrt{\frac{P_{i}}{3}}|\Psi\rangle_{S}|3\rangle_E,
\end{eqnarray}
and now the Kraus operators are
\begin{eqnarray}
M_0^{i}=\sqrt{1-P_{i}}\sigma_{0},~~~  M_1^{i}=
\sqrt{\frac{P_{i}}{3}}\sigma_{1},~~~
M_2^{i}=\sqrt{\frac{P_{i}}{3}}\sigma_{2},~~~
M_3^{i}=\sqrt{\frac{P_{i}}{3}}\sigma_{3}, \label{Eq.8}
\end{eqnarray}
where $\sigma_{\mu}$ are the Pauli operators,  $i=(1,2 \cdots N)$
and $\mu=(0,1,2,3)$.

For both two environments $P_{i}$ is the decay parameter  relating
only to time. Under the Markov approximation, the relationship
between the parameter $P_i$ and the time $t$ is shown by
$P_i=(1-e^{-\Gamma_{i} t})$ \cite{21,22}, here $\Gamma_{i}$ is the decay rate.

On the other hand, the negativity is used to measure a
bipartite system $\rho_{AB}$, which is defined as \cite{Vidal}
\begin{eqnarray}\label{Eq.9}
N_{AB}= \|\rho^{T_{\alpha}}_{AB}\|-1,
\end{eqnarray}
where $T_{\alpha}$ denotes the partial transpose of $\rho_{AB}$ and
$\|.\|$ is the trace norm of a matrix. For any 3-qubit states
$|\Phi\rangle_{ABC}$,  $N_{AB}$ is  two-tangle which is the
negativity of the mixed state
$\rho_{AB}=Tr_{C}(|\Phi\rangle_{ABC}\langle\Phi|)$, and $N_{A(BC)}$
is one-tangle which is defined as
\begin{eqnarray}\label{Eq.10}
N_{A(BC)}=\|\rho^{T_{\alpha}}_{ABC}\|-1.
\end{eqnarray}
Then the so-called residual entanglement becomes
\begin{eqnarray}\label{Eq.11}
\pi_{A}=N_{A(BC)}^{2}-N_{AB}^{2}-N_{AC}^{2},
\end{eqnarray}
and the $\pi$-tangle $\pi_{ABC}$ is defined as
\begin{eqnarray}\label{Eq.12}
\pi_{ABC}=\frac{1}{3}(\pi_{A}+\pi_{B}+\pi_{C}),
\end{eqnarray}
which describes an average residual entanglement.

\section{TRIPARTITE ENTANGLEMENT for $GHZ$ initial STATE UNDER ENVIRONMENT}

We assume that Alice, Bob and Charlie share a $GHZ$ initial state
\begin{eqnarray}\label{initial}
|\Phi\rangle_{ABC}=\frac{1}{\sqrt{2}}(|0_{\omega_a}\rangle_{A} |0_{\omega_b}\rangle_{B}|0_{\omega_c}\rangle_{C}
+|1_{\omega_a}\rangle_{A}|1_{\omega_b}\rangle_{B}|1_{\omega_c}\rangle_{C}),
\end{eqnarray}
where $|0_{\omega_{a(b,c)}}\rangle_{A(B,C)}$ and
$|1_{\omega_{a(b,c)}}\rangle_{A(B,C)}$ are vacuum states and the
first excited states from the perspective of an inertial observer.
We also assume that Alice, Bob and Charlie each carry a monochromatic
detector sensitive to frequencies $\omega_a$, $\omega_b$ and
$\omega_c$, respectively. Using Eq. (\ref{Eq.1}) and tracing over
the disconnected region $II$, we can get the state in Rindler
spacetime
\begin{eqnarray}\label{Eq.14}
|\Phi\rangle_{AB_{I}C_{I}}&=&\frac{1}{\sqrt{2}}[\cos{r_{b}}
\cos{r_{c}}|0\rangle_{A}|0\rangle_{B_{I}}|0\rangle_{C_{I}}+
\cos{r_{b}}\sin{r_{c}}|0\rangle_{A}|0\rangle_{B_{I}}
|1\rangle_{C_{I}}\nonumber
\\&&+\sin{r_{b}}\cos{r_{c}}|0\rangle_{A} |1\rangle_{B_{I}}
|0\rangle_{C_{I}}+
\sin{r_{b}}\sin{r_{c}}|0\rangle_{A}|1\rangle_{B_{I}}
|1\rangle_{C_{I}}+|1\rangle_{A}|1\rangle_{B_{I}}|1\rangle_{C_{I}}],
\end{eqnarray}
hereafter frequency subscripts are dropped. Then we obtain the
density matrix
\begin{eqnarray}\label{Eq.15}
\rho_{AB_{I}C_{I}}&=&\frac{1}{2}[\cos^{2}{r_{b}}\cos^{2}{r_{c}}
|000\rangle\langle000|+
\cos^{2}{r_{b}}\sin^{2}{r_{c}}|001\rangle\langle001|\nonumber
\\&&+\sin^{2}{r_{b}}\cos^{2}{r_{c}}|010\rangle\langle010|+
\sin^{2}{r_{b}}\sin^{2}{r_{c}}|011\rangle\langle011|\nonumber
\\&&+\cos{r_{b}}\cos{r_{c}}(|000\rangle\langle111|+
|111\rangle\langle000|)+|111\rangle\langle111|],
\end{eqnarray}
where $|mnl\rangle=|m\rangle_{A}|n\rangle_{B_{I}}
|l\rangle_{C_{I}}$. For simplification, in what follows we just
consider the case that both Bob and Charlie move with the same
acceleration, i. e. $r_{b}=r_{c}=r$.

\subsection{Amplitude damping channel}

Now we let all the subsystems interact with amplitude damping
environment.  Using Eqs. (\ref{Eq.3}) and (\ref{Eq.4}), we get the
evolved state
\begin{eqnarray}\label{Eq.16}
\rho^{evo}_{AB_{I}C_{I}}&=&\frac{1}{2}\{[\cos^{4}{r}+(n+m)
\cos^{2}{r}\sin^{2}{r}+mn\sin^{4}{r}+p m n]|000\rangle\langle000
|\nonumber
\\&&+
[(1-n)(\cos^{2}{r}\sin^{2}{r}+m\sin^{4}{r}+pm)]
|001\rangle\langle001|+(1-p)(1-m)n|110\rangle\langle110|\nonumber
\\&&+[(1-m)(\cos^{2}{r}\sin^{2}{r}+n\sin^{4}{r}+p n)]
|010\rangle\langle010|+(1-p)(1-n)m|101\rangle\langle101|\nonumber
\\&&+[(1-m)(1-n)(\sin^{4}{r}+p)]|011\rangle\langle011
|+(1-p)m n|100\rangle\langle100|\nonumber
\\&&+
\sqrt{(1-p)(1-m)(1-n)}\cos^{2}{r}(|000\rangle\langle111|+
|111\rangle\langle000|)\nonumber
\\&&+(1-p)(1-m)(1-n)|111\rangle\langle111|\},
\end{eqnarray}
where $p$, $m$, $n$ are the decay probability for Alice, Bob and
Charlie, respectively. After some calculations we find one-tangles
\begin{eqnarray}  \label{Eq.17}
N_{A(B_{I}C_{I})}&=&\frac{1}{2}\{-1-m n-p+m p+n p+\cos^{4}{r}
+2\cos^{2}{r}\sin^{2}{r}+m\sin^{4}{r}+n\sin^{4}{r}-m n\sin^{4}{r}
   \nonumber
\\&&+\sqrt{(-1+p)[m^{2} n^{2}(-1+p)-(-1+m)(-1+n)\cos^{4}{r}]}
   \nonumber
\\&&+
   \sqrt{(-1 + m) (-1 +
      n) [-(-1+p)\cos^{4}{r}+(-1+m)(-1+n)(p+\sin^{4}{r})^{2}]}\}.
\end{eqnarray}
\begin{eqnarray}  \label{Eq.18}
N_{B_{I}(AC_{I})}&=&\frac{1}{2}\{\sqrt{(-1 + m)(-(-1+p)(-1+n)
\cos^{4}{r}+(-1+m)[\cos^{2}{r}\sin^{2}{r}+n(p+\sin^{4}{r})]^{2})}
\nonumber
\\&&+\sqrt{(-1+n)(-1+p)[m^{2} (-1+n)(-1+p)-(-1+m)\cos^{4}{r}]}
   +\sin^{4}{r}+m n\sin^{4}{r}\nonumber
\\&& -1-m+m n+m p-n p+\cos^{4}{r}+\cos^{2}{r}\sin^{2}{r}+m\cos^{2}
{r}\sin^{2}{r}-n\sin^{4}{r}
   \},
\end{eqnarray}
\begin{eqnarray}  \label{Eq.19}
N_{C_{I}(AB_{I})}&=&\frac{1}{2}\{\sqrt{(-1+n)(-(-1+p)(-1+m)
\cos^{4}{r}+(-1+n)[\cos^{2}{r}\sin^{2}{r}+m(p+\sin^{4}{r})]^{2})}
\nonumber
\\&&+\sqrt{(-1+m)(-1+p)[n^{2} (-1+m)(-1+p)-(-1+n)\cos^{4}{r}]}
   +\sin^{4}{r}+m n\sin^{4}{r}\nonumber
\\&& -1-n+m n-m p+n p+\cos^{4}{r}+\cos^{2}{r}\sin^{2}{r}+
n\cos^{2}{r}\sin^{2}{r}-m\sin^{4}{r}
   \}.
\end{eqnarray}

\begin{figure}[ht]
\includegraphics[scale=0.42]{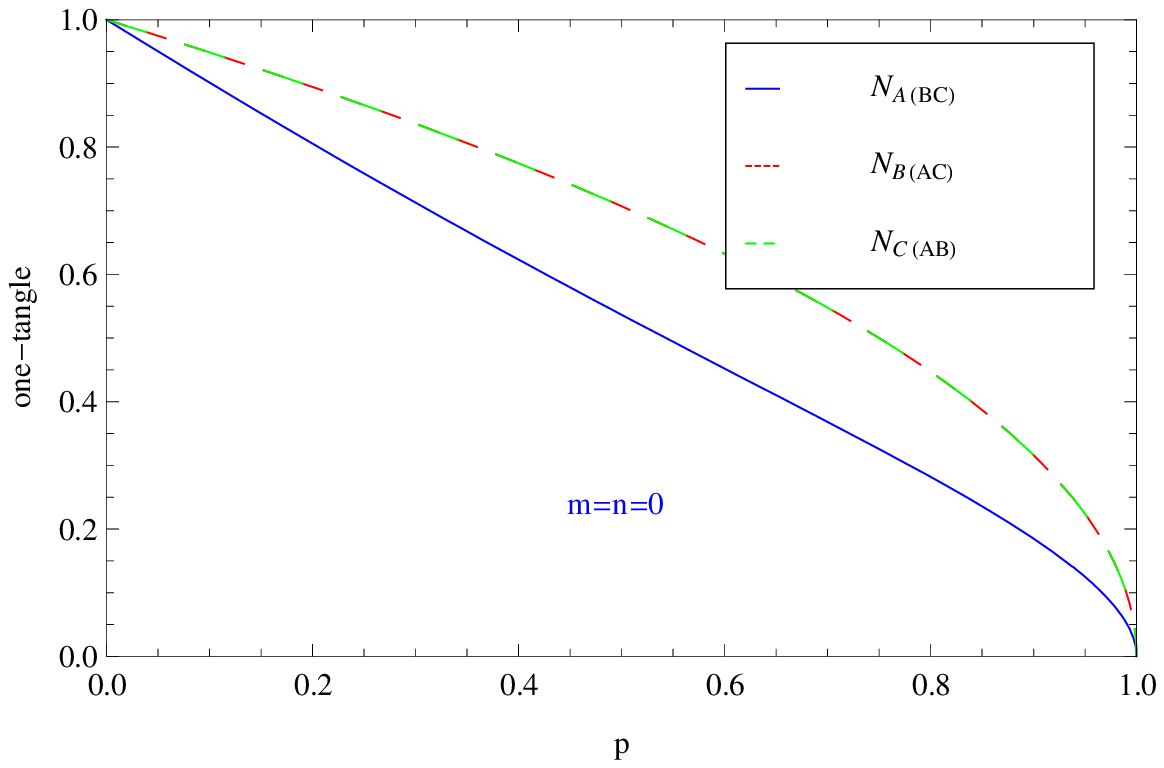}
\includegraphics[scale=0.42]{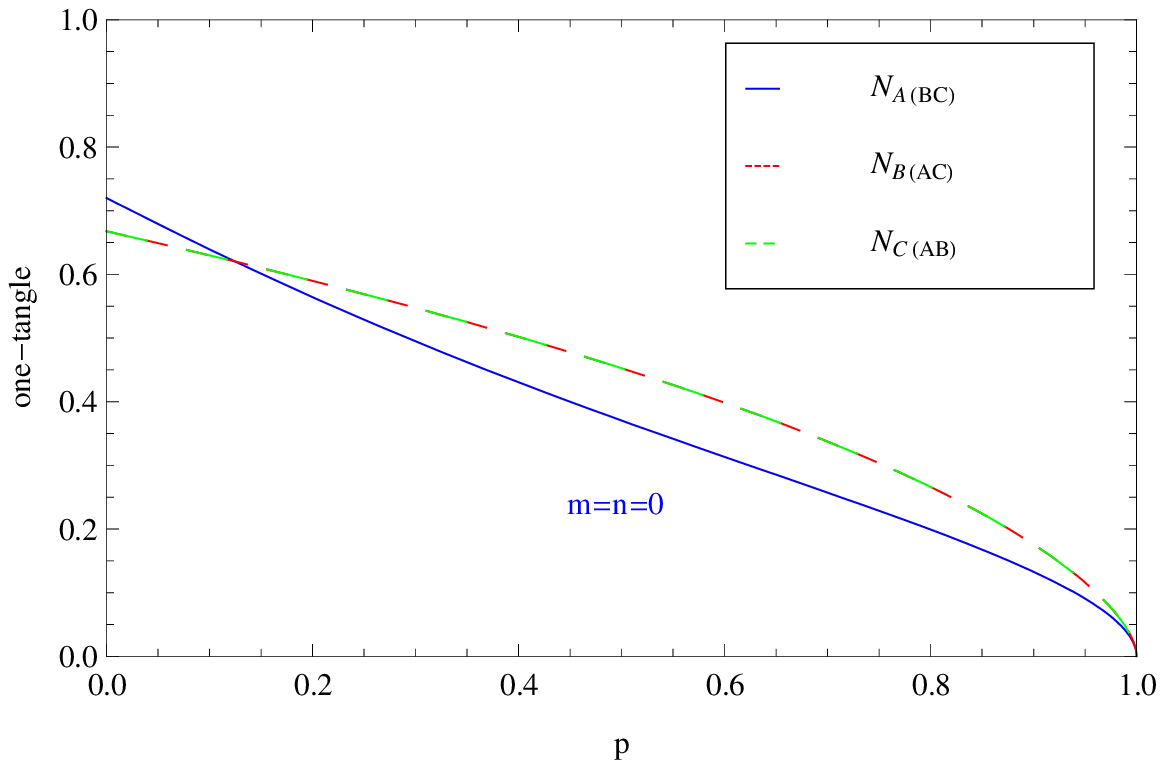}
\includegraphics[scale=0.42]{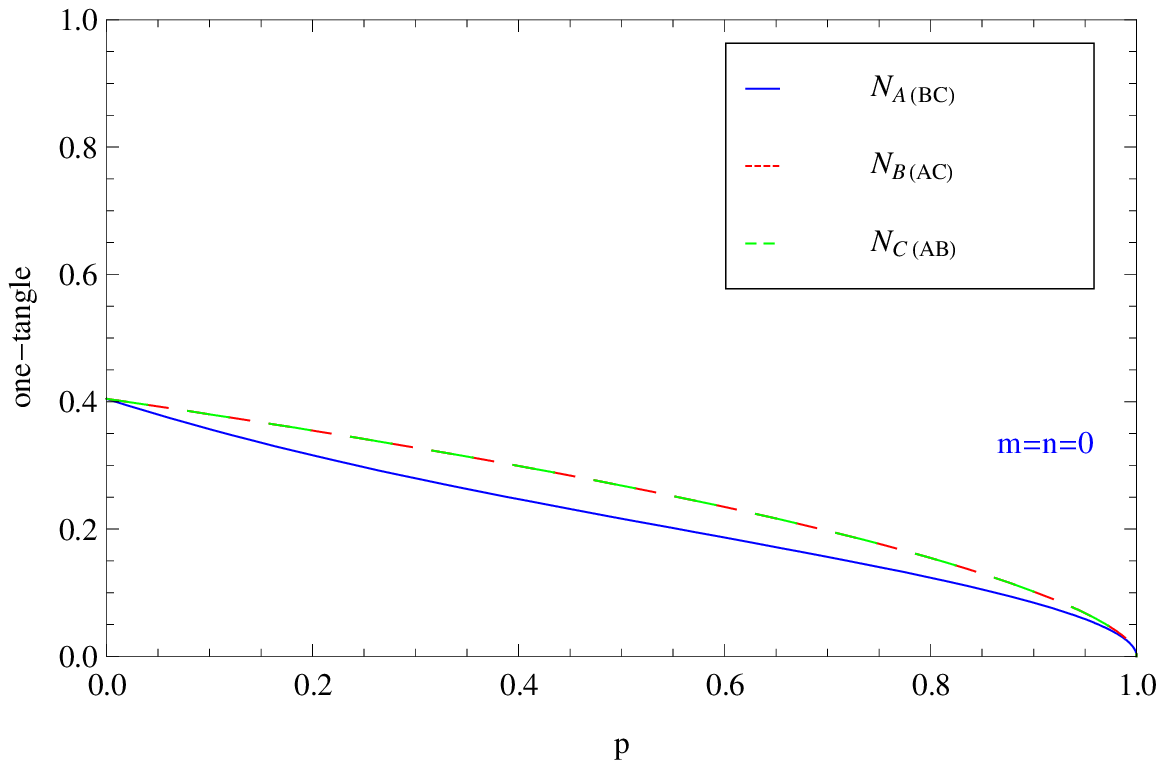}\\
\includegraphics[scale=0.43]{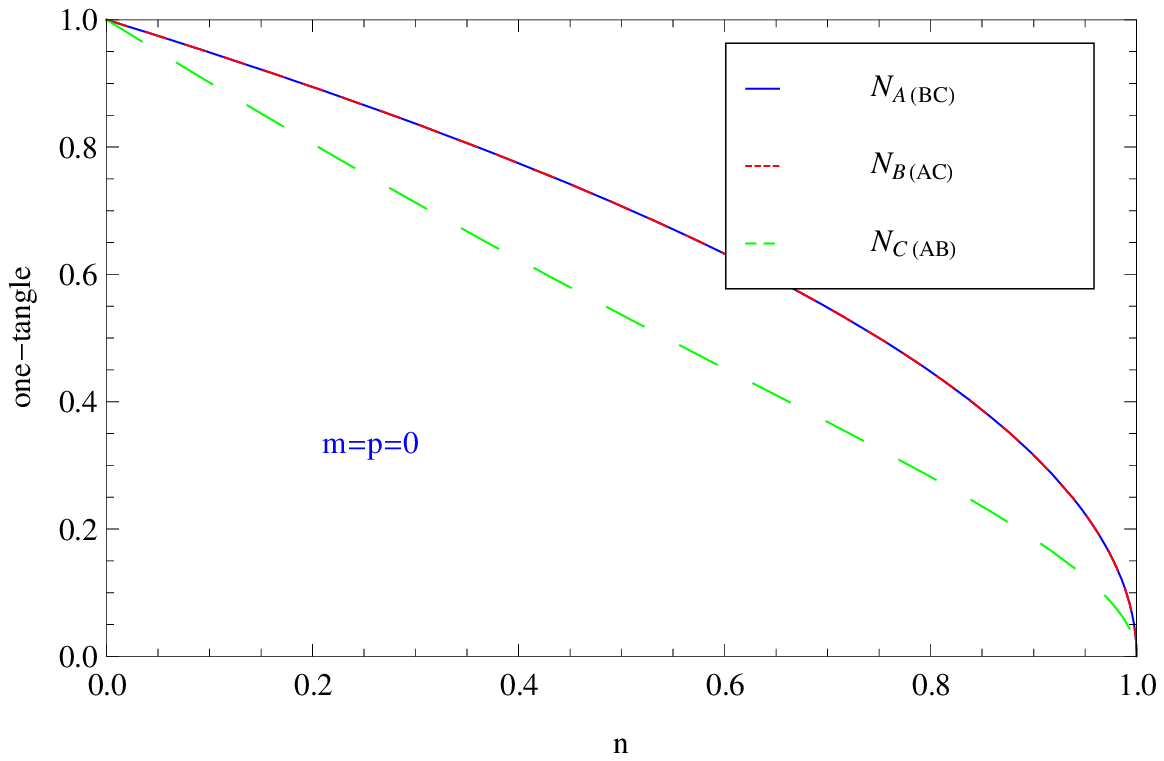}
\includegraphics[scale=0.42]{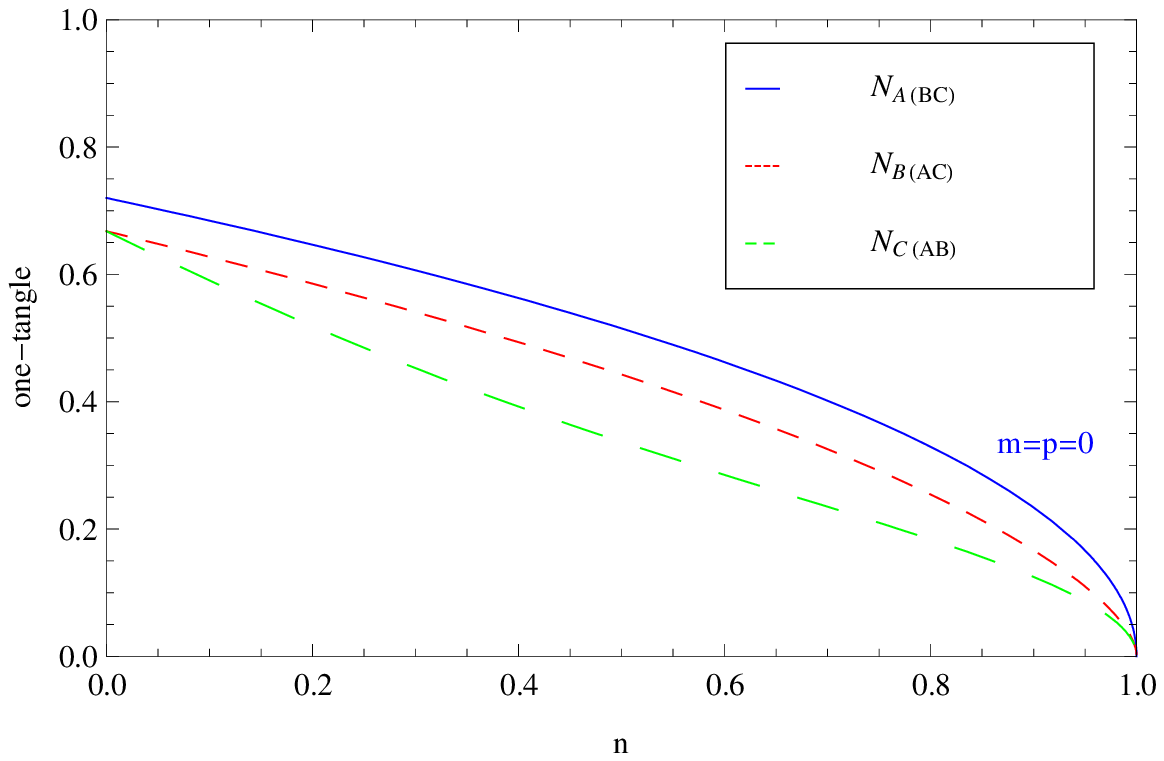}
\includegraphics[scale=0.42]{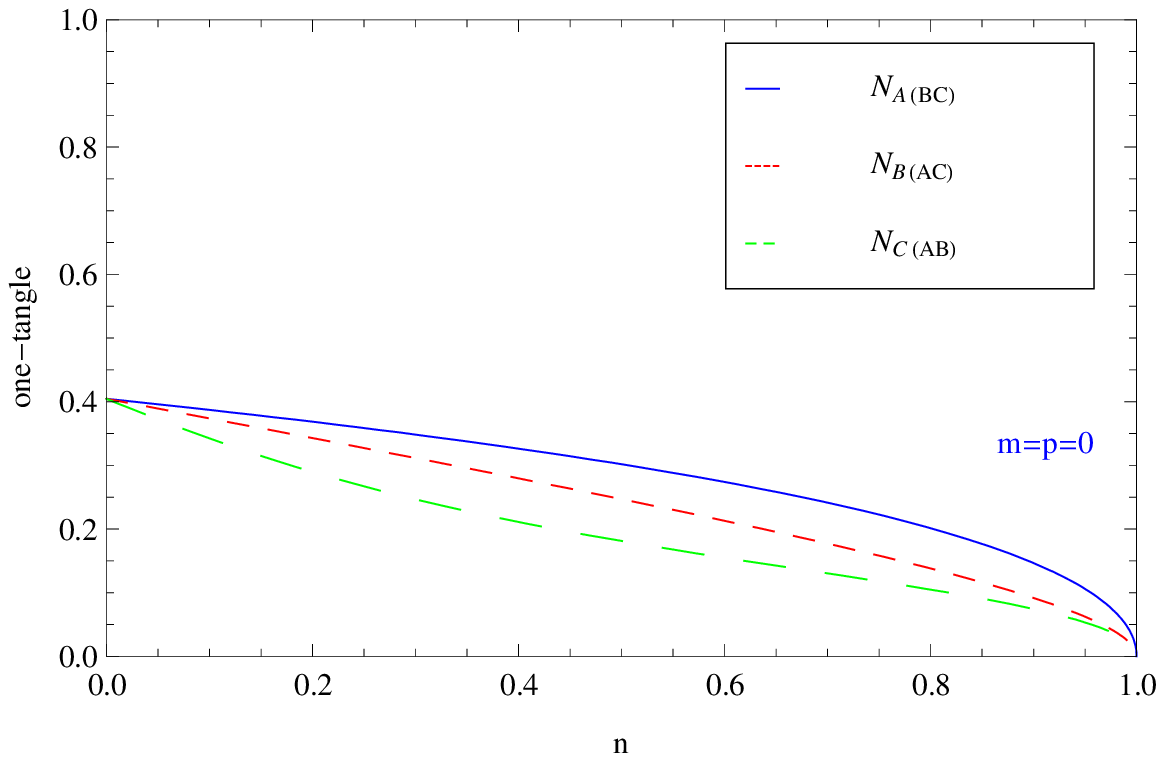}\\
\includegraphics[scale=0.43]{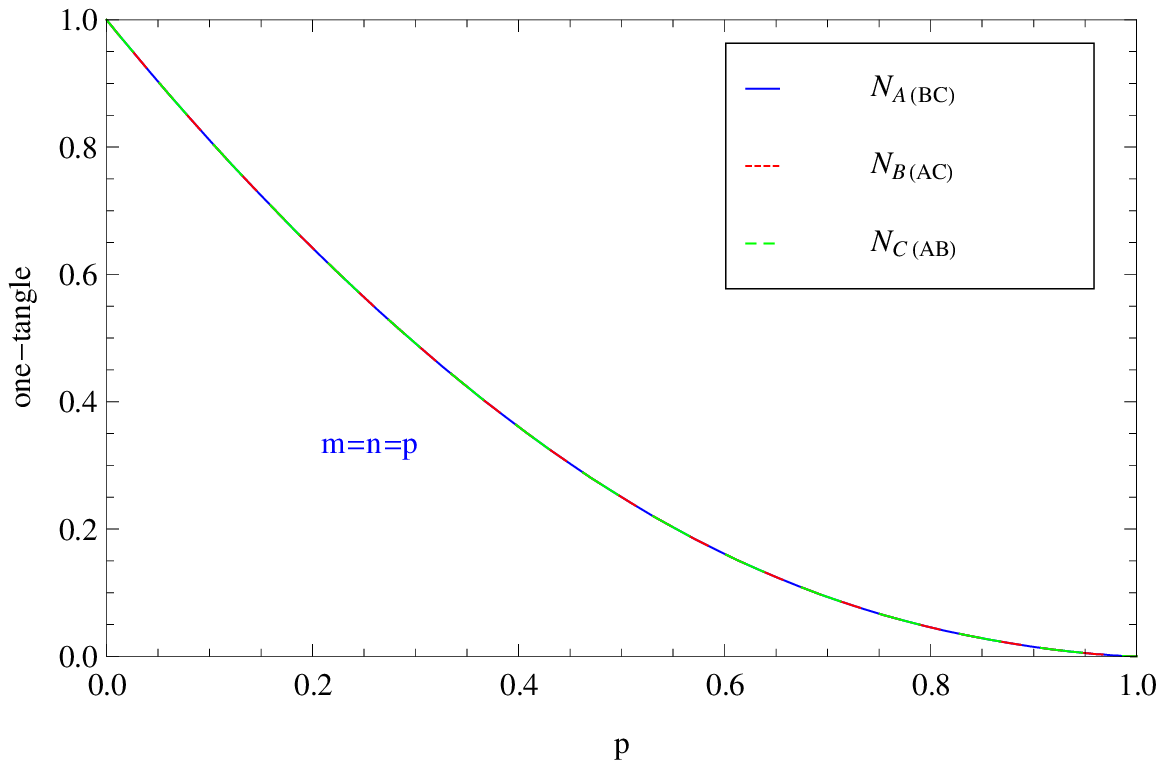}
\includegraphics[scale=0.42]{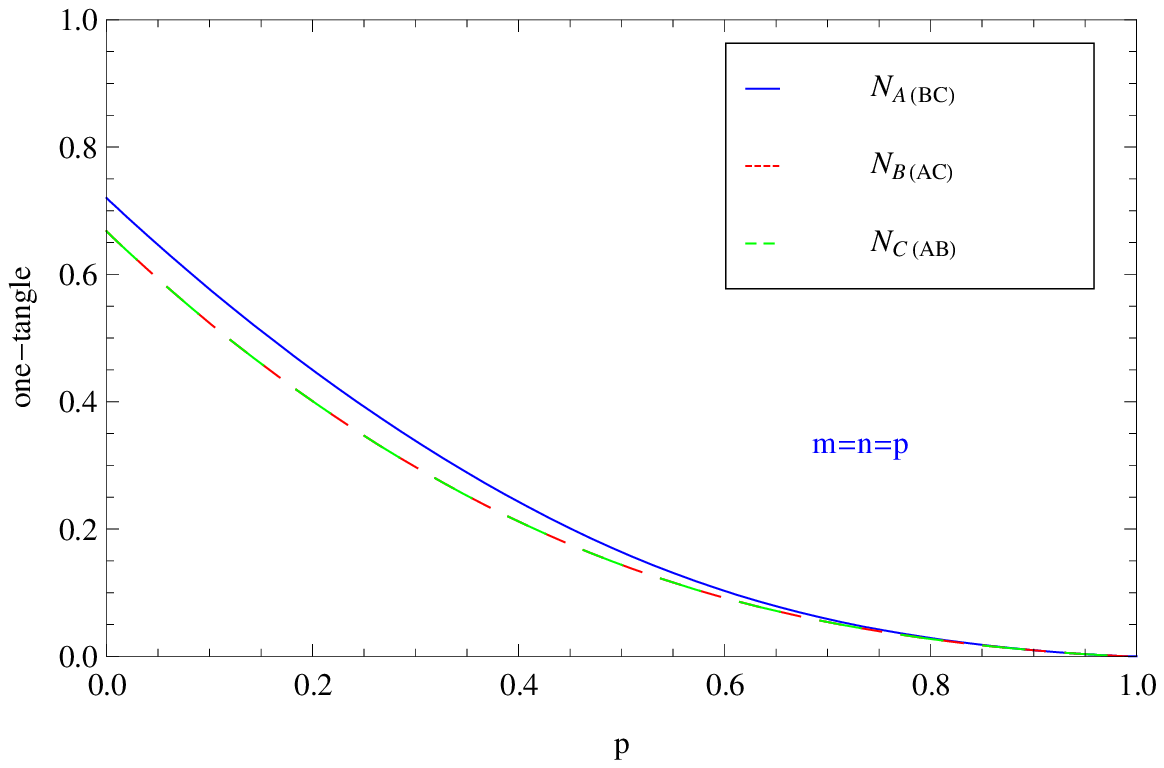}
\includegraphics[scale=0.42]{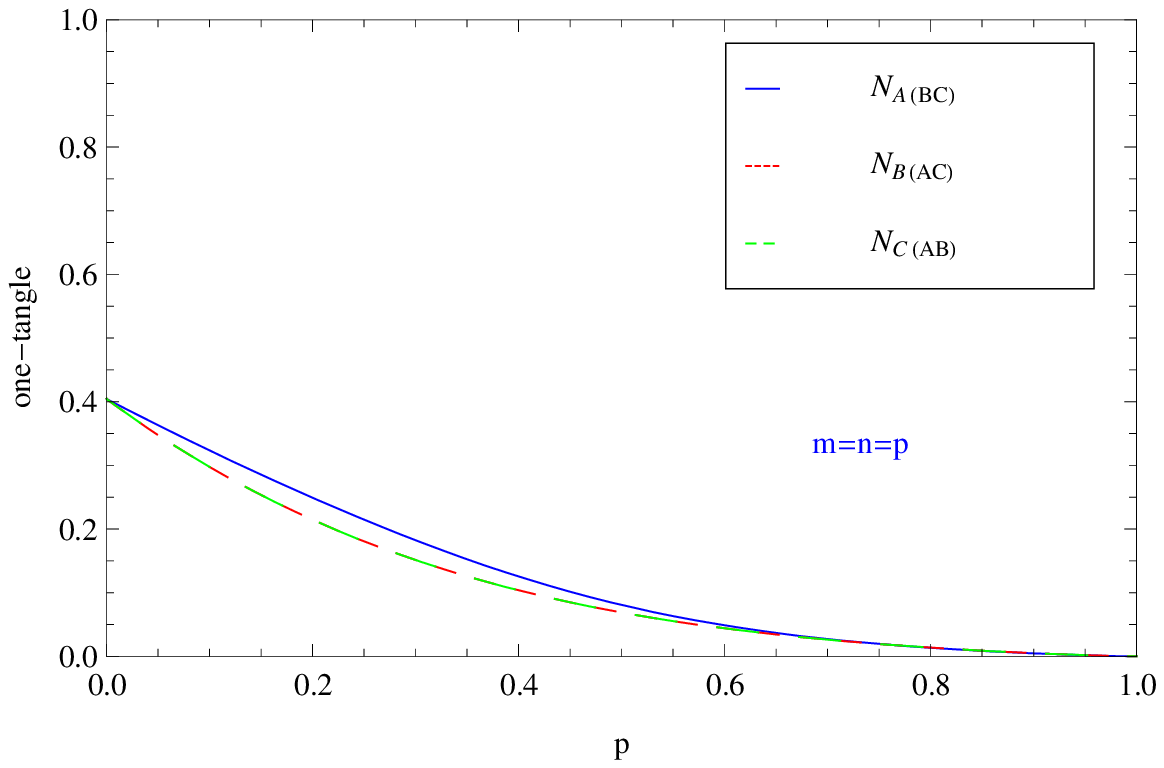}
\caption{\label{Fig.1}(Color online) The plot shows the negativity
$N_{A(B_{I}C_{I})}$ (blue line ), $N_{B_{I}(AC_{I})}$ (red line) and
$N_{C_{I}(AB_{I})}$ (green line) for amplitude damping channel. The
first (second) row presents that only inertial observer Alice
(noninertial observer Charlie) is under the environment. And the
third row is for the case that Alice, Bob, and Charlie all interact
with the environment. We draw them for $r=0$ (left rank), $r=\pi/6$
(middle rank), and $r=\pi/4$ (right rank).}
\end{figure}

One-tangles are shown by the first row in Fig.\ref{Fig.1} with
$m=n=0$, which means that only the inertial observer Alice interacts
with environment. All the one-tangles decrease as the interaction
increases, and disappear completely when $p=1$ which means the entire
tripartite entanglement is destroyed at an infinite time. Note that
$N_{B_{I}(AC_{I})}=N_{C_{I}(AB_{I})}$ for any time which indicates
Bob's and Charlie's subsystem are symmetry when only Alice acts with the
environment. The intersect point in the middle picture indicates
that $N_{A(B_{I}C_{I})}=N_{B_{I}(AC_{I})}=N_{C_{I}(AB_{I})}$, which
means that there is no difference among all the subsystems at this
point. Generally, the intersect point is
\begin{eqnarray}\label{Eq.20}
p&=&\cos{2r}\sin^{2}{r},
\end{eqnarray}
which means that when $p$ and $r$ satisfy this relationship  we
can't distinguish the three subsystems by the negativity. We also
see that a bigger acceleration means a smaller initial entanglement
for one-tangle as expected. When $r=\pi/4$ the three one-tangles
have the same initial entanglement which is the same as the result
in \cite{17}. It is worth to note that the sudden death never occurs
for one-tangle even Bob and Charlie are in the limit of infinite
acceleration.

We show the one-tangles by second row in Fig. \ref{Fig.1}  with
$p=m=0$, which means that only the noninertial observer Charlie is
under the amplitude damping environment. Obviously, if $r=0$ the
decay curve would be the same with the former case with $r=0$.
However, if $r\neq 0$ we see that $N_{B_{I}(AC_{I})}$ decreases
more slowly than $N_{C_{I}(AB_{I})}$ because Bob doesn't interact with
environment while Charlie does, and  $N_{A(B_{I}C_{I})}$ decays
more slowly than $N_{B_{I}(AC_{I})}$ because Bob has an acceleration
while Alice doesn't. That is to say, both the acceleration and
environment can destroy the symmetry between the subsystems, which
can be used to distinguish them. It is interesting to note that no
intersect point and no sudden death exist in this case.

The situation with $m=n=p$ is shown by the third row in Fig.
\ref{Fig.1}, which means Alice, Bob, and Charlie are all under the same
environment. It is easy to find out that all the one-tangles
decrease more quickly than the former two cases, which is similar to
the behaviors of bipartite entanglement. Especially, if $r=0$ we
find all the three subsystems have the same decay curve due to they
are highly symmetric and indistinguishable. And if $r\neq 0$,
$N_{B_{I}(AC_{I})}=N_{C_{I}(AB_{I})}$ for all the time because of
their symmetry again. Even all the subsystems interact with the
environment and in the infinite acceleration there is still no
sudden death yet.

On the other hand, by use of Eq. (\ref{Eq.9}) we compute the
two-tangle between any two subsystems of the multipartite system.
Tracing the qubit of $C_{I}$ we get $N_{AB_{I}}=0$, which means that
no bipartite entanglement exist between $A$ and $B_{I}$ with
considering of both environment and acceleration. Similarly, it is
easy to obtain $N_{AC_{I}}=N_{B_{I}C_{I}}=0$.
\begin{figure}[ht]
\includegraphics[scale=0.43]{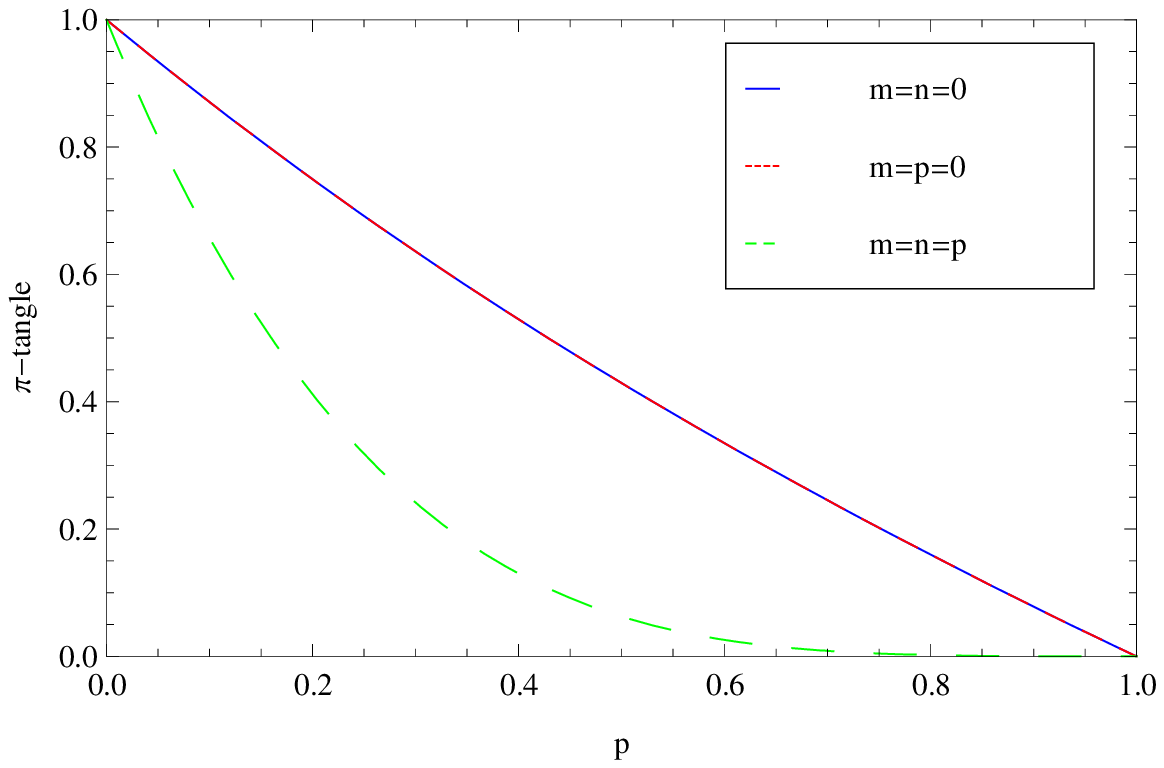}
\includegraphics[scale=0.42]{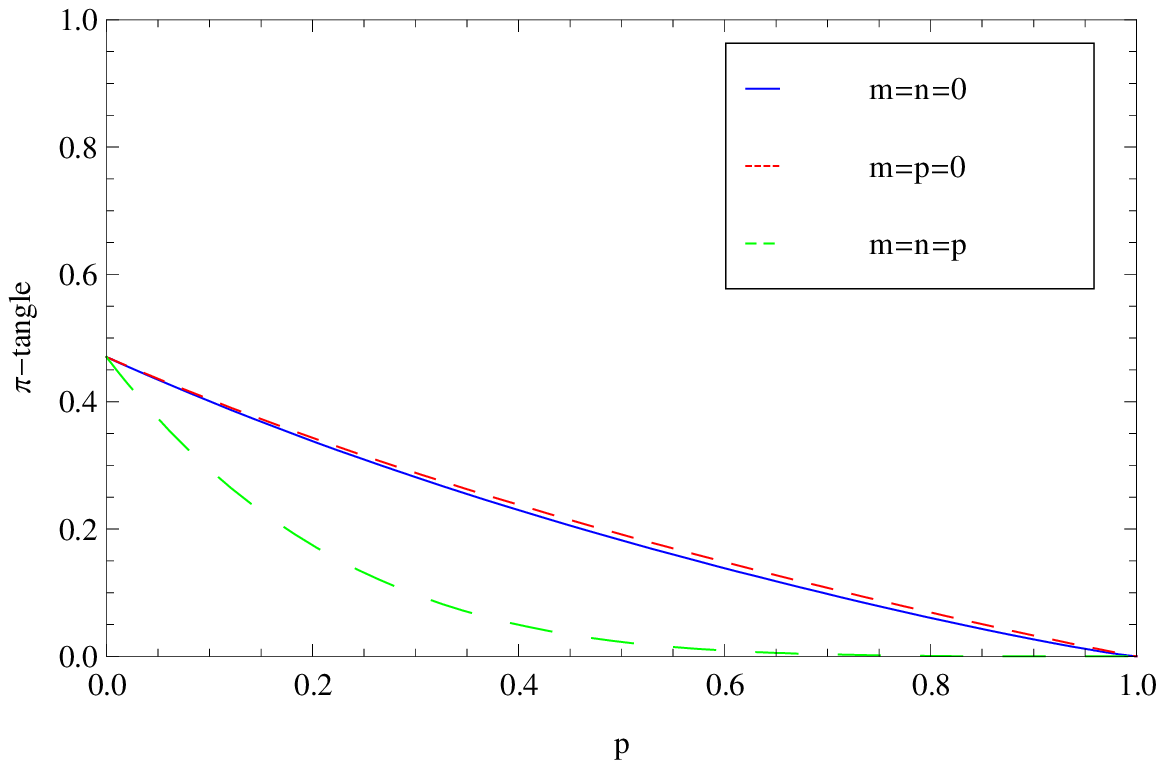}
\includegraphics[scale=0.42]{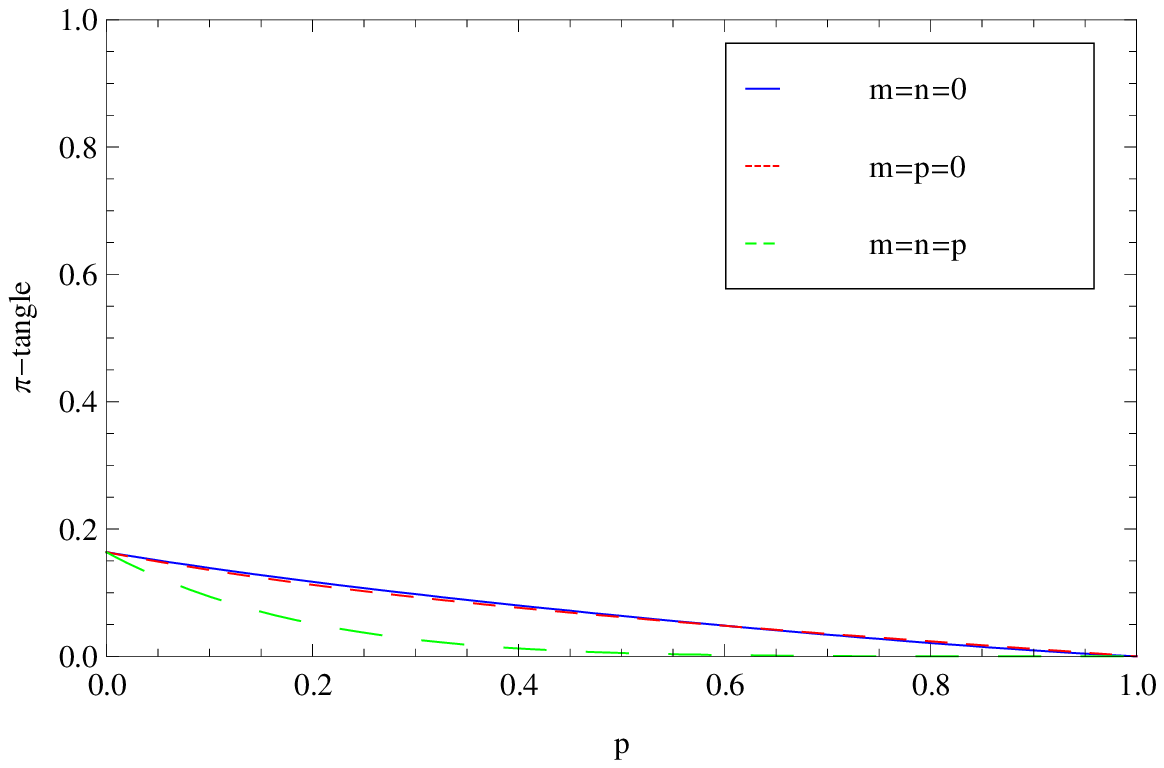}
\caption{\label{Fig.2}(Color online) Blue (red) line plots the
$\pi$-tangle for the case that only Alice (Charlie) interacts with
the environment. Green line plots the case of all of them
interacting with the environment. We show three cases for $r=0$
(left), $r=\pi/6$ (middle), and $r=\pi/4$ (right). It is worth to
note that there is still no sudden death.}
\end{figure}

In addition, we calculate the $\pi$-tangle by use of Eqs.
(\ref{Eq.11}) and (\ref{Eq.12})
\begin{eqnarray}\label{Eq.21}
\pi_{AB_{I}C_{I}}&=&\frac{1}{3}(\pi_{A}+\pi_{B_{I}}+\pi_{C_{I}})
=\frac{1}{3}[N_{A(B_{I}C_{I})}^{2}+N_{B_{I}(AC_{I})}^{2}+N_{C_{I}
(AB_{I})}^{2}].
\end{eqnarray}
We give the results in Fig. \ref{Fig.2} for the former three cases.
Note that a bigger acceleration also means a smaller initial
$\pi$-tangle just like before. It is interesting to note that the
$\pi$-tangle decay curves for $m=n=0$ and $m=p=0$ are almost the
same, which indicates that the effect of environment is so strong
that we can nearly ignore the effect of acceleration for tripartite
entanglement. We also find that the more strong subsystems interact
with environment, the more quickly the $\pi$-tangle decreases. We
can prove that there is still no sudden death yet. Taking the
highest possible case when $m=n=p, r=\pi/4$, we have
\begin{eqnarray}\label{Eq.22}
&&\pi_{AB_{I}C_{I}}=\frac{1}{192}\{2[1 + 4 p  -2\sqrt{(p-1
)^{3}(-1-4p^{2} + 4
p^{3})}-(1-p)\sqrt{5-2p+9p^{2}+8p^{3}+16p^{4}}]^{2}\nonumber
\\&&+[3 p^{2}-1 - 2 p+\sqrt{(p-1 )^{3} (-5 - 7 p -
8 p^2 + 16 p^{3})} -2 (p-1) \sqrt{1 - p + 4 p^4}]^{2}- 5 p^{2}
 \} .
\end{eqnarray}
It is easy for us to find that $\pi_{AB_{I}C_{I}}$ is monotone
decreasing function when $0<p<1$ and it exactly equals to zero when
$p=1$ which indicates that no sudden death appears.

\subsection{Depolarizing noise}

\begin{figure}[ht]
\includegraphics[scale=0.43]{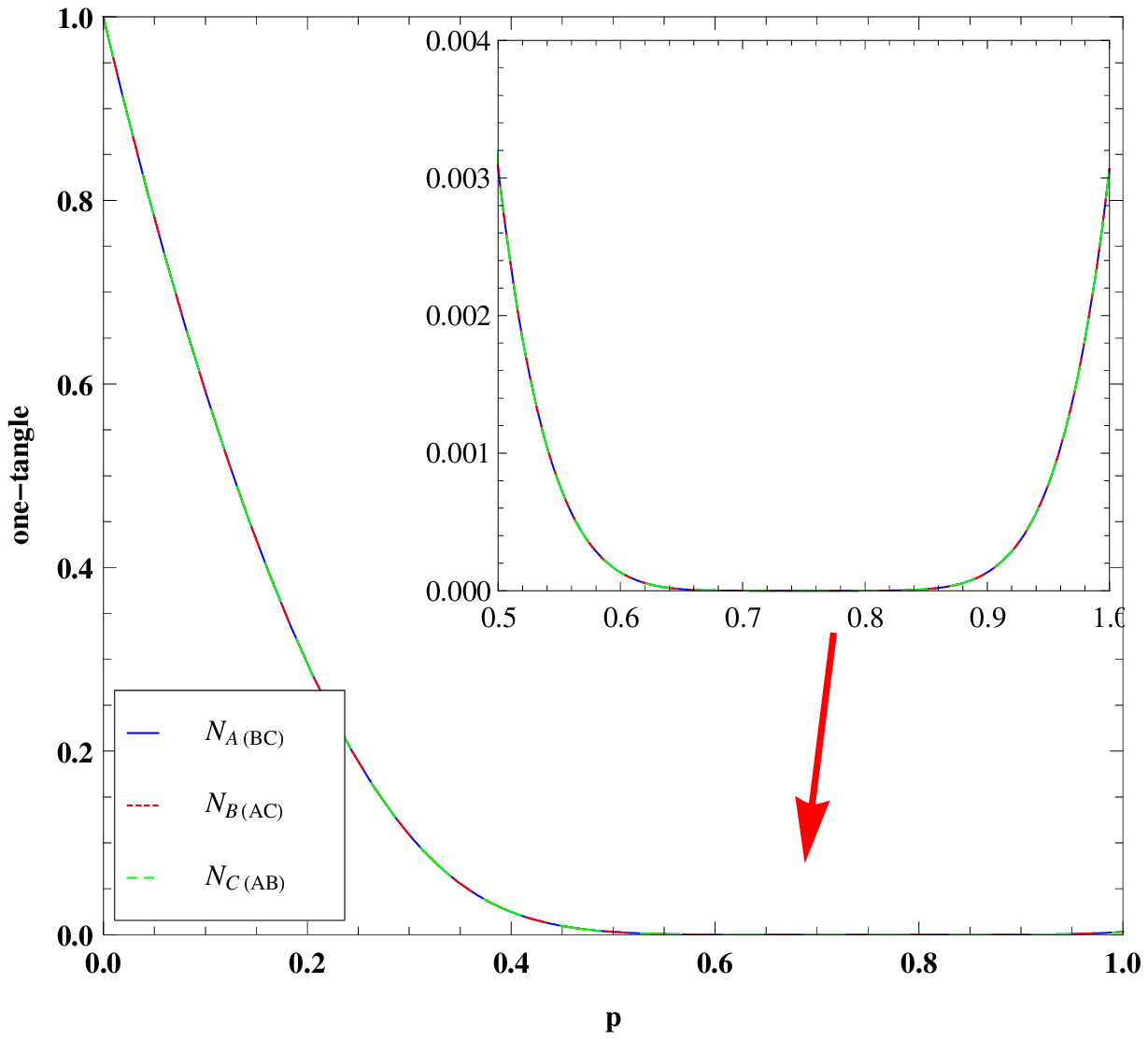}
\includegraphics[scale=0.42]{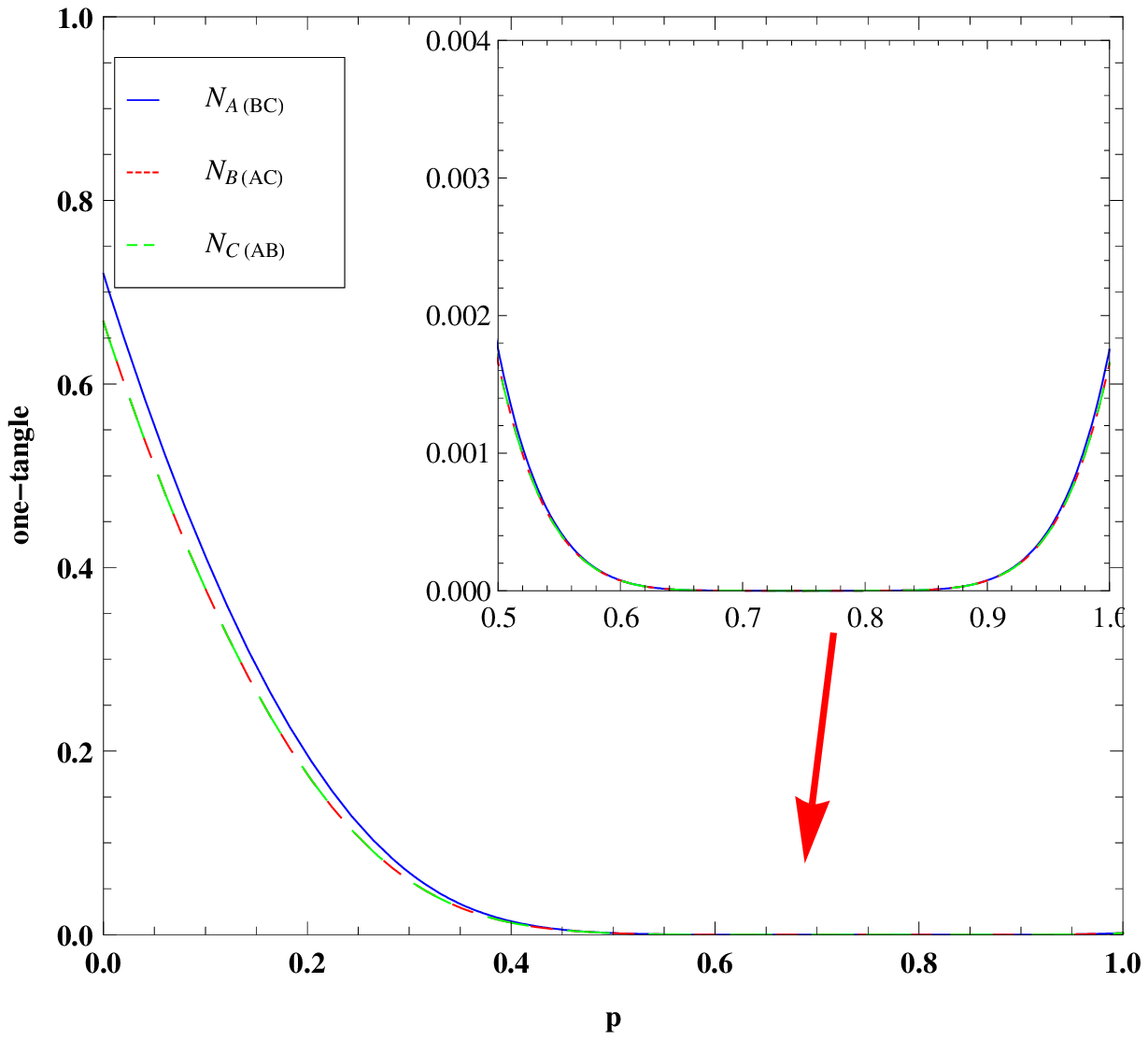}
\includegraphics[scale=0.42]{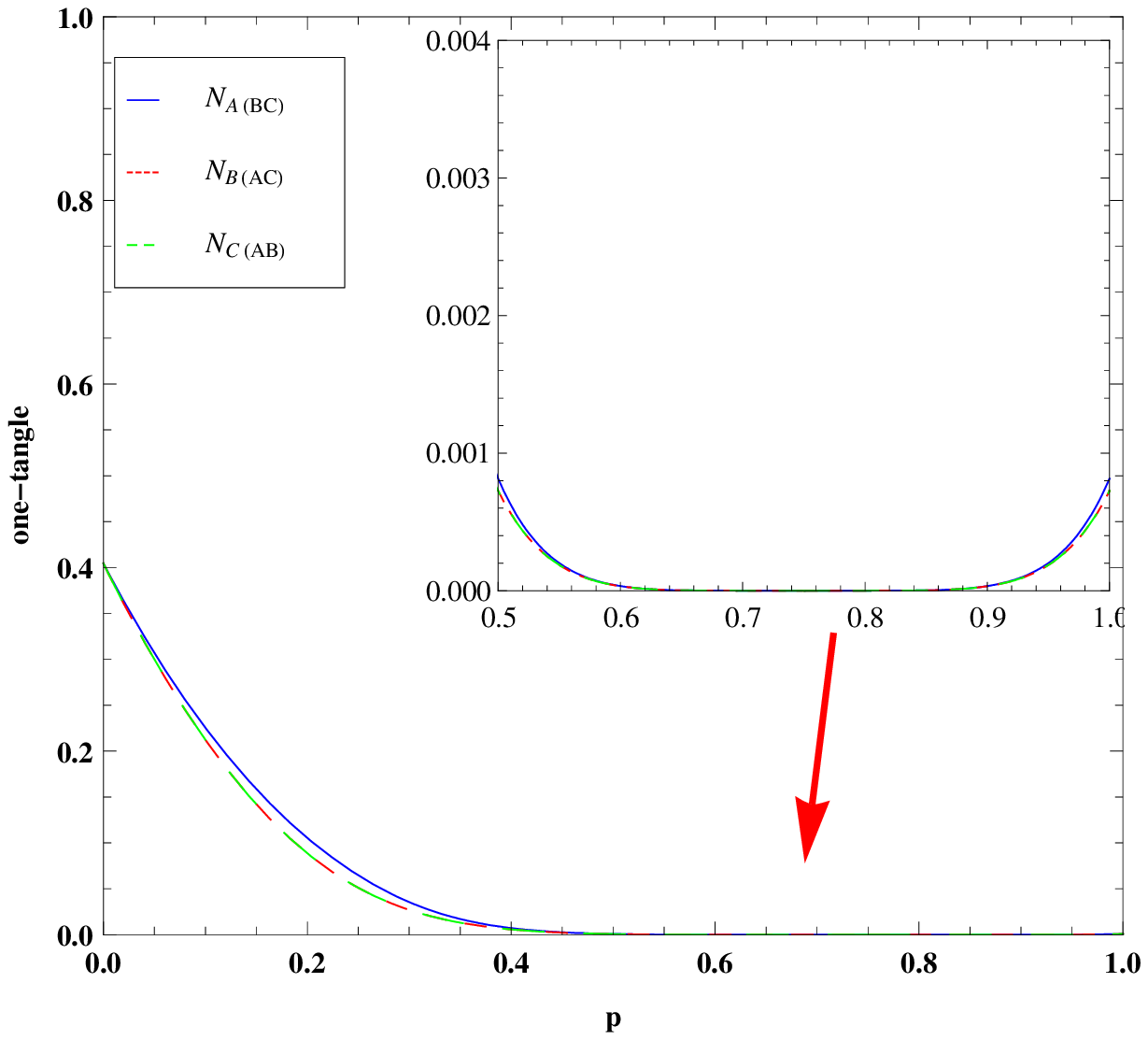}
\caption{\label{Fig.3}(Color online) The negativity
$N_{A(B_{I}C_{I})}$ (blue line ), $N_{B_{I}(AC_{I}})$ (red line),
and $N_{C_{I}(AB_{I})}$ (green line) when all Alice, Bob, and
Charlie are in depolarizing noise. We show three cases for $r=0$
(left), $r=\pi/6$ (middle), and $r=\pi/4$ (right). The rebound
process for entanglement is plotted in the magnifying pictures.}
\end{figure}

We present the results in Fig. \ref{Fig.3} for the case that all the
subsystems are in depolarizing noise. We see that many
characteristics for amplitude damping channel still remain under
this environment. But the one-tangle in depolarizing noise decays
much more quickly than it goes under amplitude damping channel.
What's surprising is that the one-tangles  decays to zero at
$p=0.75$ and then a rebound process takes place when $p>0.75$. This
means that all the tripartite entanglement transfers to environment
at $p=0.75$ and then part of it transfers from environment back to
the system when $p>0.75$. The bigger the acceleration is, the
smaller this rebound process becomes.

\begin{figure}[ht]
\includegraphics[scale=0.45]{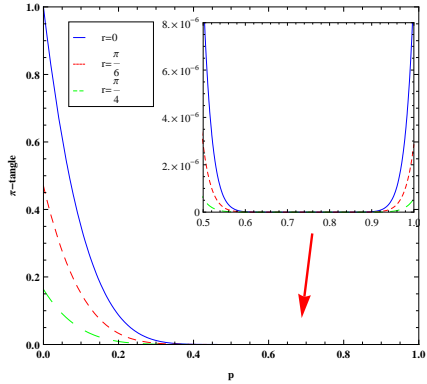}
\caption{\label{Fig.4}(Color online) The $\pi$-tangle
$\pi_{AB_{I}C_{I}}$ when Alice, Bob, and Charlie are all in
depolarizing noise. We plot three cases for $r=0$ (blue line ),
$r=\pi/6$ (red line), and $r=\pi/4$ (green line). The rebound
process for entanglement is plotted in the magnifying picture.}
\end{figure}

In addition, using Eq. (\ref{Eq.9}) we get again that
$N_{AB_{I}}=N_{AC_{I}}=N_{B_{I}C_{I}}=0$, which is exactly the same
as before. That is to say, either in amplitude damping channel or in
depolarizing noise, either in inertial frame or in noninertial
frame, there is no bipartite entanglement in this system for $GHZ$
state. We also get the $\pi$-tangle by use of Eqs. (\ref{Eq.11}) and
(\ref{Eq.12}), then plot it and its rebound process in
Fig.\ref{Fig.4} .

At last, we note that the CKW inequality \cite{23}, $N_{AB}^{2}
+N_{AC}^{2}\leq N_{A(BC)}^{2}$, is saturated for this initial state,
which means the effect of both environment and acceleration doesn't
destroy this inequality for $GHZ$ initial state.

\section{TRIPARTITE ENTANGLEMENT for $W$ STATE
 UNDER THE ENVIRONMENT  }

Now we assume  Alice, Bob and Charlie share a $W$ initial state
\begin{eqnarray}\label{Eq.23}
|\Phi\rangle_{ABC}=\frac{1}{\sqrt{3}}(|0\rangle_{A}|0\rangle_{B}
|1\rangle_{C}+|0\rangle_{A}|1\rangle_{B}|0\rangle_{C}+|1\rangle_{A}
|0\rangle_{B}|0\rangle_{C}).
\end{eqnarray}
With the help of  Eq. (\ref{Eq.1}) we obtain the system's density
matrix
\begin{eqnarray}\label{Eq.24}
\rho_{AB_{I}C_{I}}&&=\frac{1}{3}[\cos^{2}{r_{b}}|001\rangle\langle
001|+
\cos^{2}{r_{c}}|010\rangle\langle010|+(\sin^{2}{r_{b}}+\sin^{2}
{r_{c}})|011\rangle\langle011|\nonumber
\\&&+
\cos^{2}{r_{b}}\cos^{2}{r_{c}}|100\rangle\langle100|+\cos^{2}
{r_{b}}\sin^{2}{r_{c}}|101\rangle\langle101|
+\cos^{2}{r_{c}}\sin^{2}{r_{b}}|110\rangle\langle110|\nonumber
\\&&+\sin^{2}{r_{b}}\sin^{2}{r_{c}}|111\rangle\langle111|
+\cos{r_{b}}\cos{r_{c}}(|010\rangle\langle001|+|001\rangle
\langle011|) \nonumber
\\&&+\cos^{2}{r_{b}}\cos{r_{c}}(|100\rangle\langle001|+|001
\rangle\langle100|)+\cos{r_{b}}\cos^{2}{r_{c}}(|100\rangle
\langle010|+|010\rangle\langle100|)\nonumber
\\&&
+\cos{r_{b}}\sin^{2}{r_{c}}(|101\rangle\langle011|+|011\rangle
\langle101|)+\cos{r_{c}}\sin^{2}{r_{b}}(|110\rangle\langle011|
+|011\rangle\langle110|)].
\end{eqnarray}

\subsection{Amplitude damping channel}

Here we also just consider Bob and Charlie move with the same
acceleration, i. e. $r_{b}=r_{c}=r$. In amplitude damping channel,
by use of Eqs. (\ref{Eq.3}) and (\ref{Eq.4}), then we get the
evolved state
\begin{eqnarray}\label{Eq.25}
\rho^{evo}_{AB_{I}C_{I}}&=&\frac{1}{3}\{[(m+n)\cos^{2}{r}(1+p\sin^{2}{r})+p\cos^{4}{r}+mn\sin^{2}{r}(2+p\sin^{2}{r})]|000\rangle\langle000|\nonumber
\\&&+\frac{1}{8}(1-n)[4+8m+p+3mp-4(-1+m(2+p))\cos{2 r}+(-1+m)p\cos{4 r}]|001\rangle\langle001|\nonumber
\\&&+\frac{1}{8}(1-m)[4+8n+p+3np-4(-1+n(2+p))\cos{2 r}+(-1+n)p\cos{4 r}]|010\rangle\langle010|\nonumber
\\&&-\frac{1}{2}(-1+m)(-1+n)(-4-p+p\cos{2 r})\sin^{2}{r}|011\rangle\langle011|\nonumber
\\&&-\frac{1}{4}(-1+p)(-1-m+(-1+m)\cos{2 r})(-1-n+(-1+n)\cos{2 r})|100\rangle\langle100|\nonumber
\\&&-\frac{1}{2}(-1+n)(-1+p)(-1-m+(-1+m)\cos{2 r})\sin^{2}{r}|101\rangle\langle101|\nonumber
\\&&-\frac{1}{2}(-1+m)(-1+p)(-1-n + (-1 + n) \cos{2 r}) \sin^{2}{r}|110\rangle\langle110|\nonumber
\\&&-(-1 + m) (-1 + n) (-1 + p) \sin^{4}{r}|111\rangle\langle111|]\nonumber
\\&&+\sqrt{1 - m} \sqrt{1 - n} \cos^{2}{r}|001\rangle\langle010|+\sqrt{1 - m} \sqrt{1 - n} \cos^{2}{r}|010\rangle\langle001|\nonumber
\\&&+\sqrt{(-1 + n) (-1 + p)}\cos{r}(\cos^{2}{r} + m \sin^{2}{r})(|100\rangle\langle001|+|001\rangle\langle100|)\nonumber
\\&&+\sqrt{(-1 + m) (-1 + p)}\cos{r}(\cos^{2}{r} + n \sin^{2}{r})(|100\rangle\langle010|+|010\rangle\langle100|)\nonumber
\\&&+\sqrt{(-1 + m) (-1 + n)^{2}(-1 + p)}\cos{r}\sin^{2}{r}(|101\rangle\langle011|+|011\rangle\langle101|)\nonumber
\\&&+\sqrt{(-1 + m)^{2} (-1 + n)(-1 + p)}\cos{r}\sin^{2}{r}(|110\rangle\langle011|+|011\rangle\langle110|)\},
\end{eqnarray}
where $p$, $m$, $n$ are the decay probability when Alice, Bob and
Charlie interact with amplitude damping environment, respectively.
Then one-tangles are given by
\begin{eqnarray}  \label{Eq.26}
N_{A(B_{I}C_{I})}&=&\frac{1}{6}\{-6 - \beta \gamma (-4 - p + p \cos{2 r}) \sin^{2}{r} -
   2 \alpha \zeta\eta +\sqrt{\beta \alpha [\beta \alpha (1 + n - \gamma \cos{2 r})^{2} \sin^{4}{r} +
4 \epsilon]}\nonumber
\\&&+ 2\sqrt{\beta \gamma \alpha \sin^{4}{r} [(-2 + m + n) \cos^{2}{r} + \beta \gamma \alpha \sin^{4}{r}]} \nonumber
\\& &+2 \sqrt{\gamma (\beta \cos^{4}{r} + \beta \gamma \alpha \cos^{2}{r} \sin^{4}{r} + \gamma [\cos^{2}{r} (1 + p \sin^{2}{r}) +m \tau\sin^{2}{r} ]^{2})} \nonumber
\\&&+ 2 \sqrt{\beta (\gamma \cos^{4}{r} + \beta \gamma \alpha \cos^{2}{r} \sin^{4}{r}
+ \beta [\cos^{2}{r} (1 + p \sin^{2}{r}) +n\tau \sin^{2}{r}]^{2})} \nonumber
\\&& +2 \sqrt{\gamma\alpha \delta+
\beta \alpha\epsilon + \varepsilon}+\sqrt{\gamma
\alpha[\gamma\alpha(1 + m - \beta \cos{2 r})^{2} \sin^{4}{r} +
4\delta]}\},
\end{eqnarray}
\begin{eqnarray}  \label{Eq.27}
N_{B(AC_{I})}&=&\frac{1}{6}\{-6 -\alpha\gamma(-1 - m + \beta \cos{2 r}) \sin^{2}{r}-2\gamma\sqrt{\beta \alpha\sin^{4}{r}(\cos^{2}{r} +\alpha\beta \sin^{4}{r})}+\nonumber
\\&&2 \sqrt{\beta\gamma\cos^{4}{r}+\alpha\beta\epsilon+[p\cos^{4}{r}+ (m+n)\cos^{2}{r}(1+p\sin^{2}{r})+m n\tau\sin^{2}{r} ]^{2}}
\nonumber
\\&&+ \sqrt{\beta \gamma [4\cos^{4}{r}+4\alpha\beta \cos^{2}{r}\sin^{4}{r}+\beta\gamma(4+p-p\cos{2 r})^{2}]}+2\sqrt{\alpha(\alpha\zeta^{2}\eta^{2}+\gamma\delta)}\nonumber
\\&& + \sqrt{\alpha\beta[4\beta\gamma\cos^{2}{r}\sin^{4}{r}+\alpha\beta(1+n-\gamma\cos{2 r})^{2}\sin^{4}{r}+4\epsilon]} \nonumber
\\&&+ 2 \sqrt{\gamma\{\alpha\beta\gamma\cos^{2}{r}\sin^{4}{r} +  \alpha\delta+\gamma[\cos^{2}{r}(1+p\sin^{2}{r})+m\tau\sin^{2}{r}]^{2}\}} \nonumber
\\&&-2\beta[\cos^{2}{r}(1+p\sin^{2}{r})+n\sin^{2}{r}(2+p\sin^{2}{r})] \},
\end{eqnarray}
\begin{eqnarray}  \label{Eq.28}
N_{C(AB_{I})}&=&\frac{1}{6}\{-6 - \alpha\beta (-1 - n + \gamma \cos{2 r}) \sin^{2}{r}-2\beta\sqrt{\gamma \alpha\sin^{4}{r}(\cos^{2}{r} +\alpha\gamma \sin^{4}{r})}+\nonumber
\\&&2 \sqrt{\beta\gamma\cos^{4}{r}+\alpha\gamma\delta+[p\cos^{4}{r}+(m+n) \cos^{2}{r}(1+p\sin^{2}{r})+m n\tau\sin^{2}{r}]^{2}}
\nonumber
\\&&+ \sqrt{\beta \gamma [4\cos^{4}{r}+4\alpha\gamma \cos^{2}{r}\sin^{4}{r}+\beta\gamma(4+p-p\cos{2 r})^{2}]}+ 2\sqrt{\alpha(\alpha\zeta^{2}\eta^{2}+\beta\delta)}\nonumber
\\&& + \sqrt{\alpha\gamma[4\beta\gamma\cos^{2}{r}\sin^{4}{r}+\alpha\gamma(1+m-\beta\cos{2 r})^{2}\sin^{4}{r}+4\delta]} \nonumber
\\&&+ 2 \sqrt{\beta\{\alpha\beta\gamma\cos^{2}{r}\sin^{4}{r} +  \alpha\epsilon+\beta[\cos^{2}{r}(1+p\sin^{2}{r})+n\tau\sin^{2}{r}]^{2}\}} \nonumber
\\&&-2\gamma[\cos^{2}{r}(1+p\sin^{2}{r})+m\sin^{2}{r}(2+p\sin^{2}{r})]
\},
\end{eqnarray}
where
\begin{eqnarray} \label{Eq.29}
\alpha&=&-1+p,~~\beta=-1+m, ~~\gamma=-1+n,~~\tau=(2+p\sin^{2}{r}),\nonumber
\\
\delta&=&(\cos{r}^{3} + m \cos{r} \sin^{2}{r})^{2},~~ \epsilon=(\cos{r}^{3} + n \cos{r} \sin^{2}{r})^{2},\nonumber
\\
\varepsilon&=&[p \cos^{4}{r} + (m + n) \cos^{2}{r} (1 + p \sin^{2}{r}) + m n \sin^{2}{r}(2 + p \sin^{2}{r})]^{2}, \nonumber
\\ \zeta&=&\cos^{2}{r} + m \sin^{2}{r},~~ \eta=\cos^{2}{r} + n \sin^{2}{r}.
\end{eqnarray}

\begin{figure}[ht]
\includegraphics[scale=0.43]{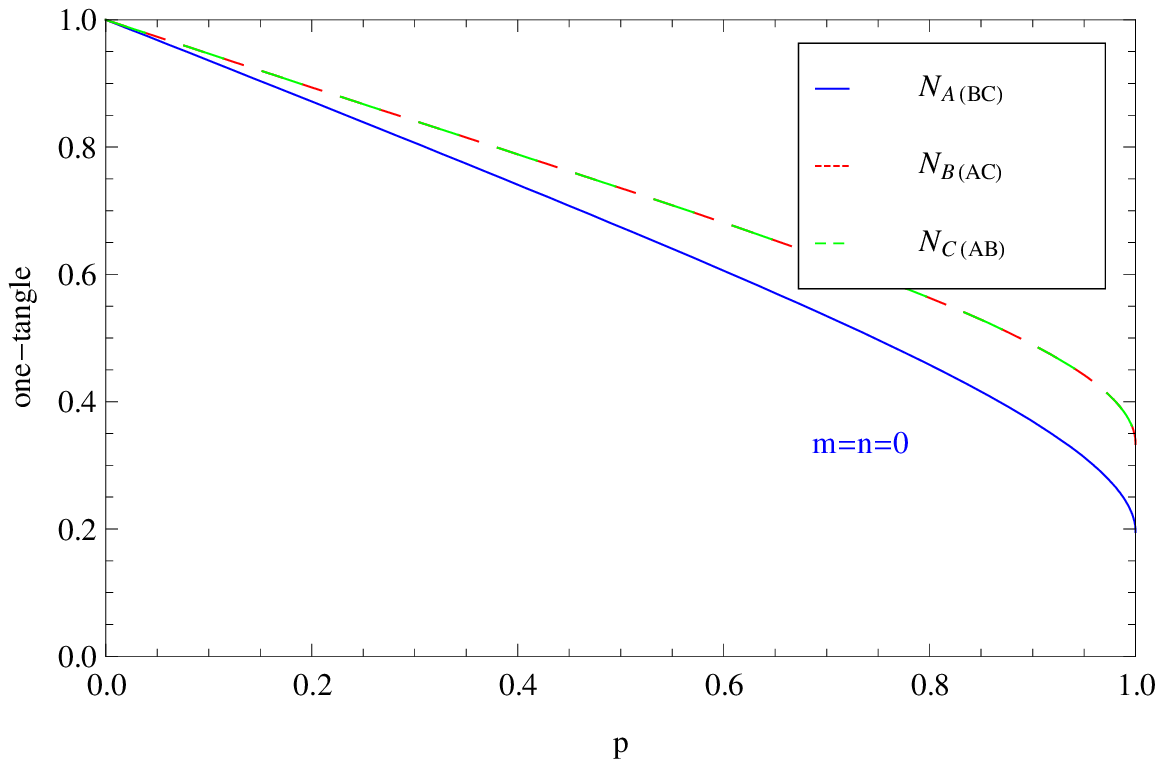}
\includegraphics[scale=0.42]{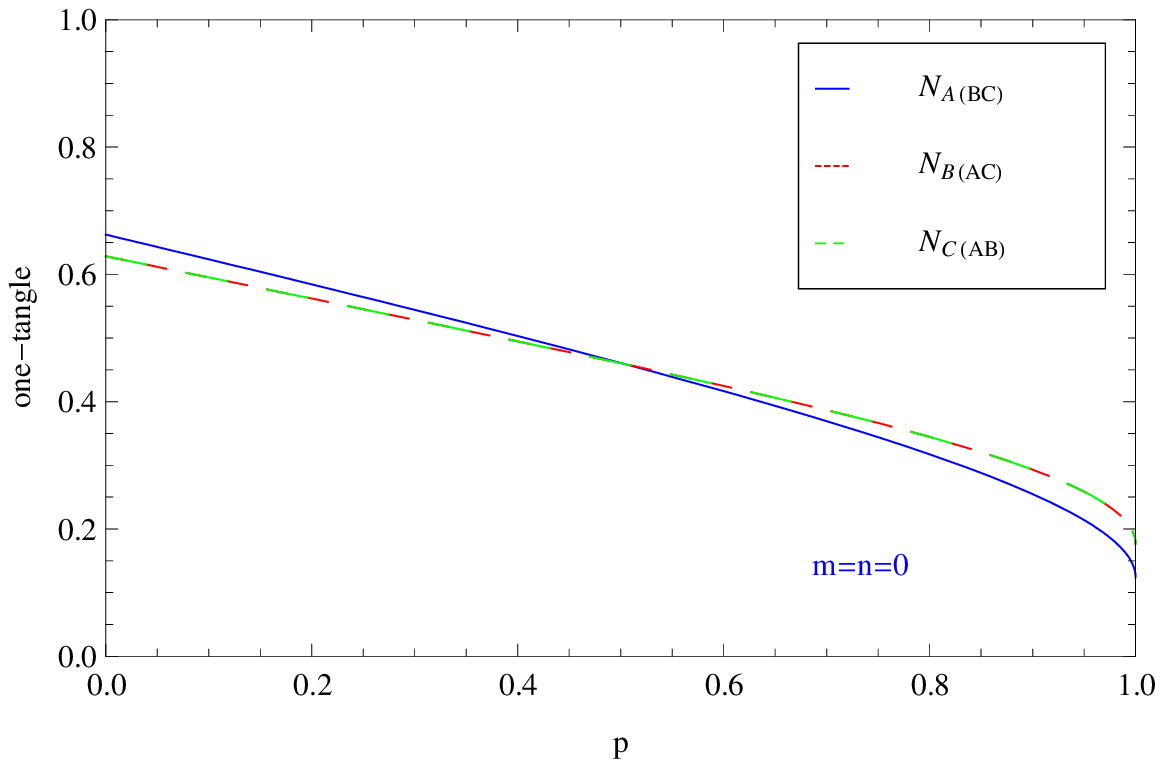}
\includegraphics[scale=0.42]{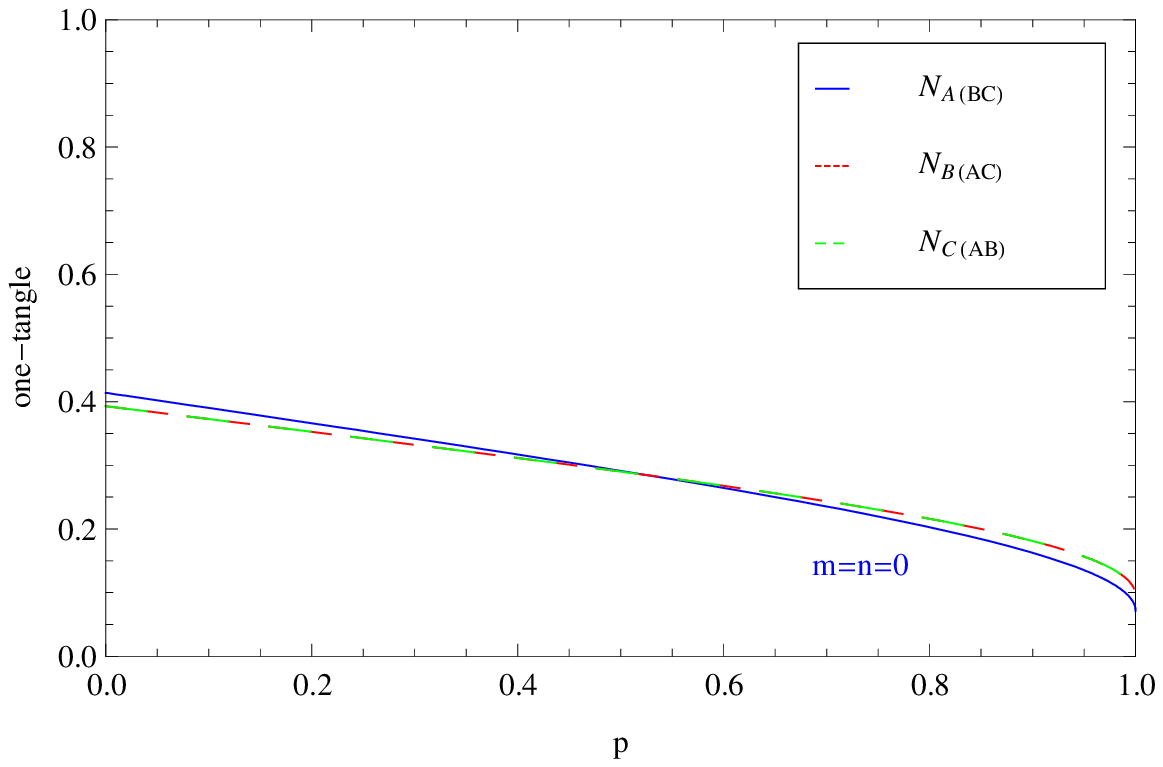}\\
\includegraphics[scale=0.43]{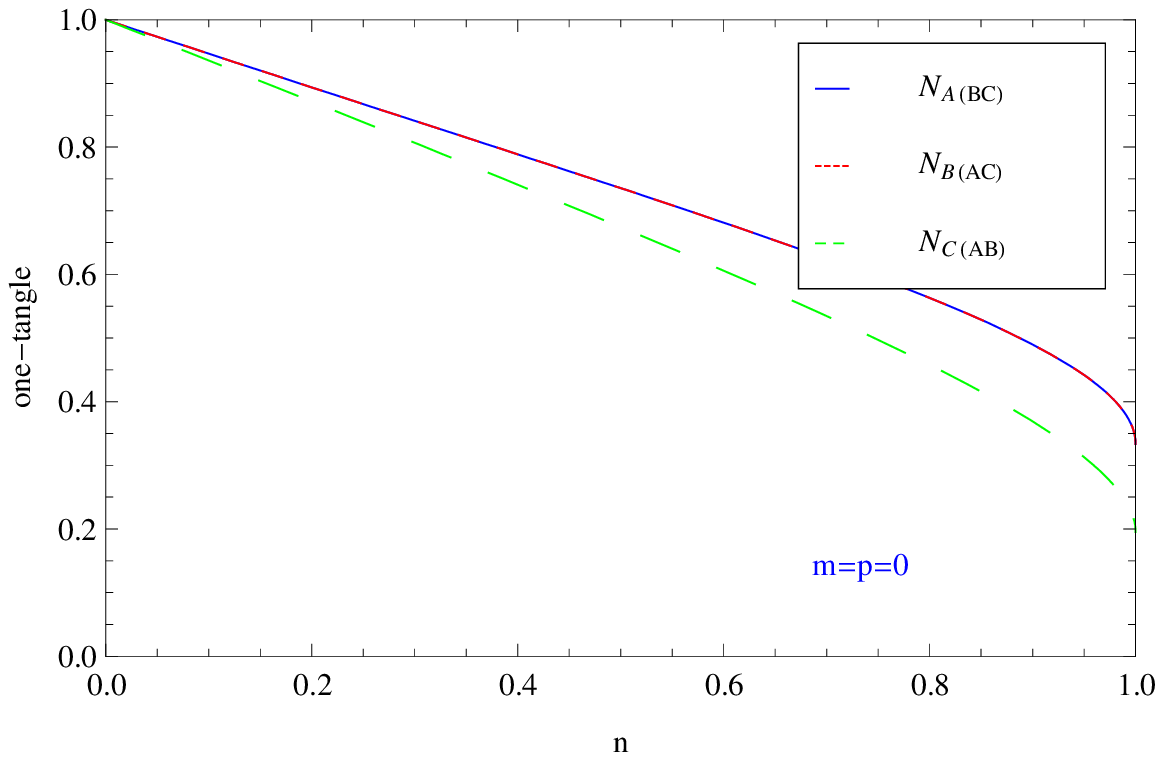}
\includegraphics[scale=0.42]{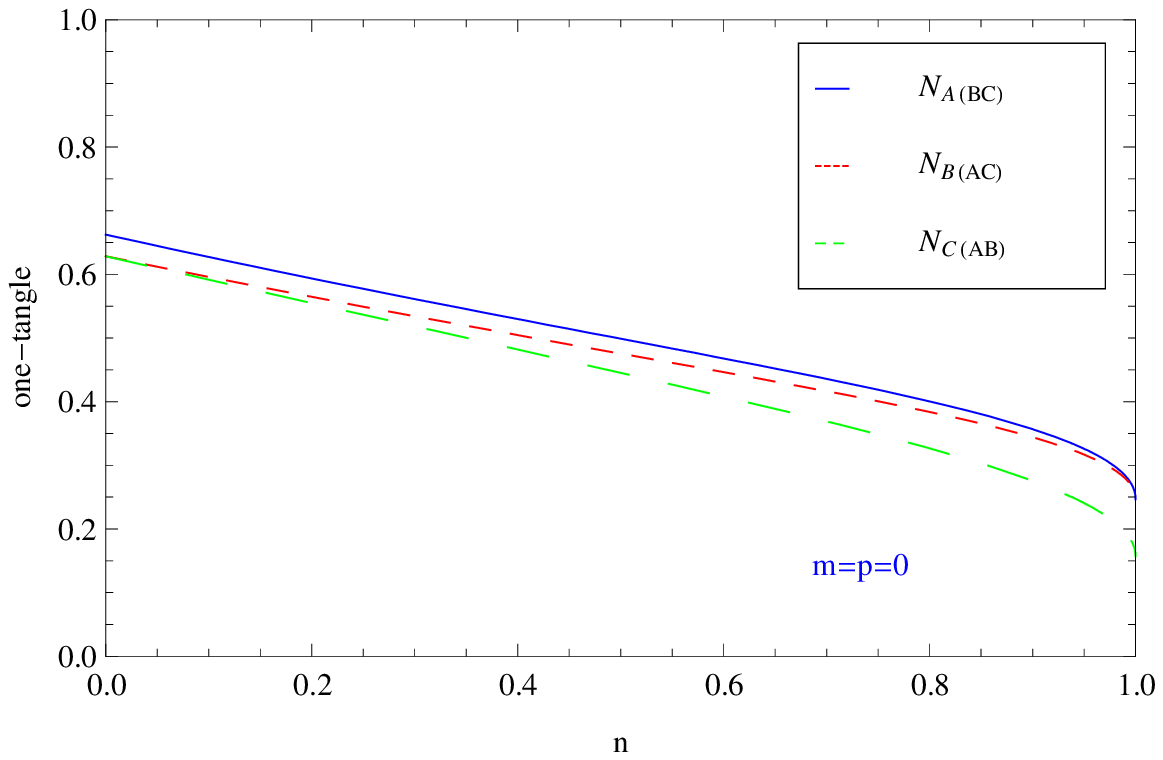}
\includegraphics[scale=0.42]{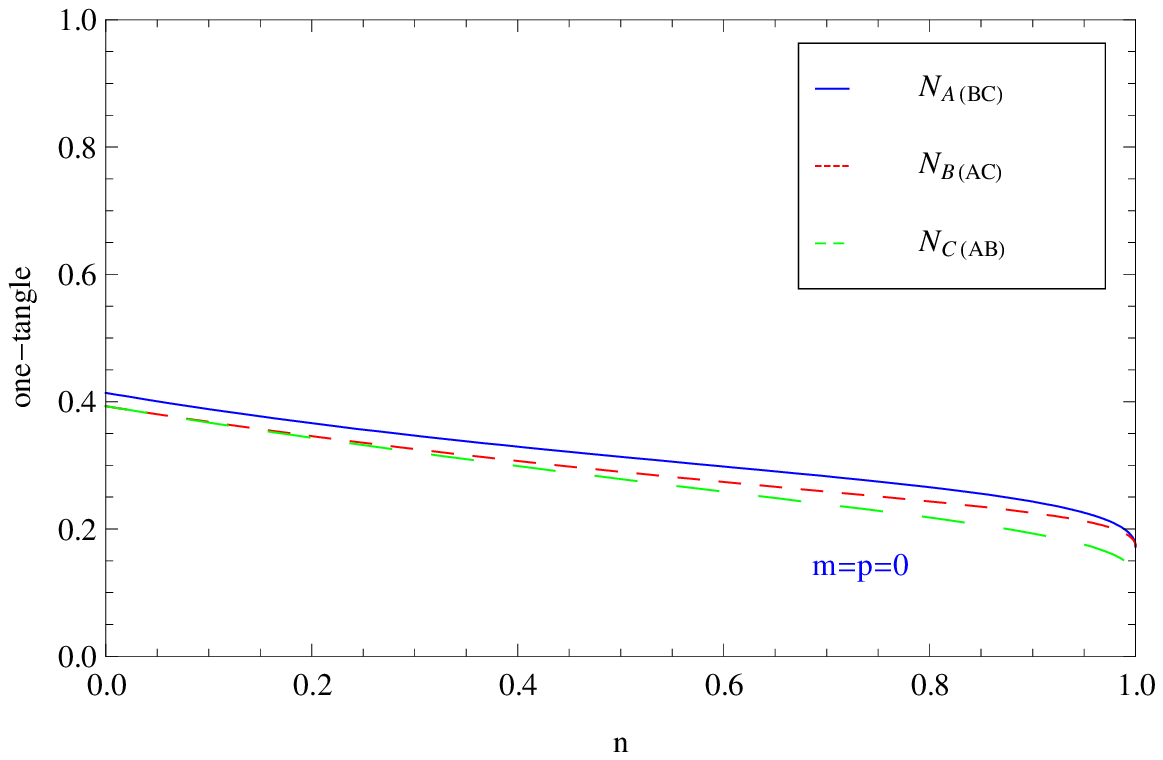}\\
\includegraphics[scale=0.43]{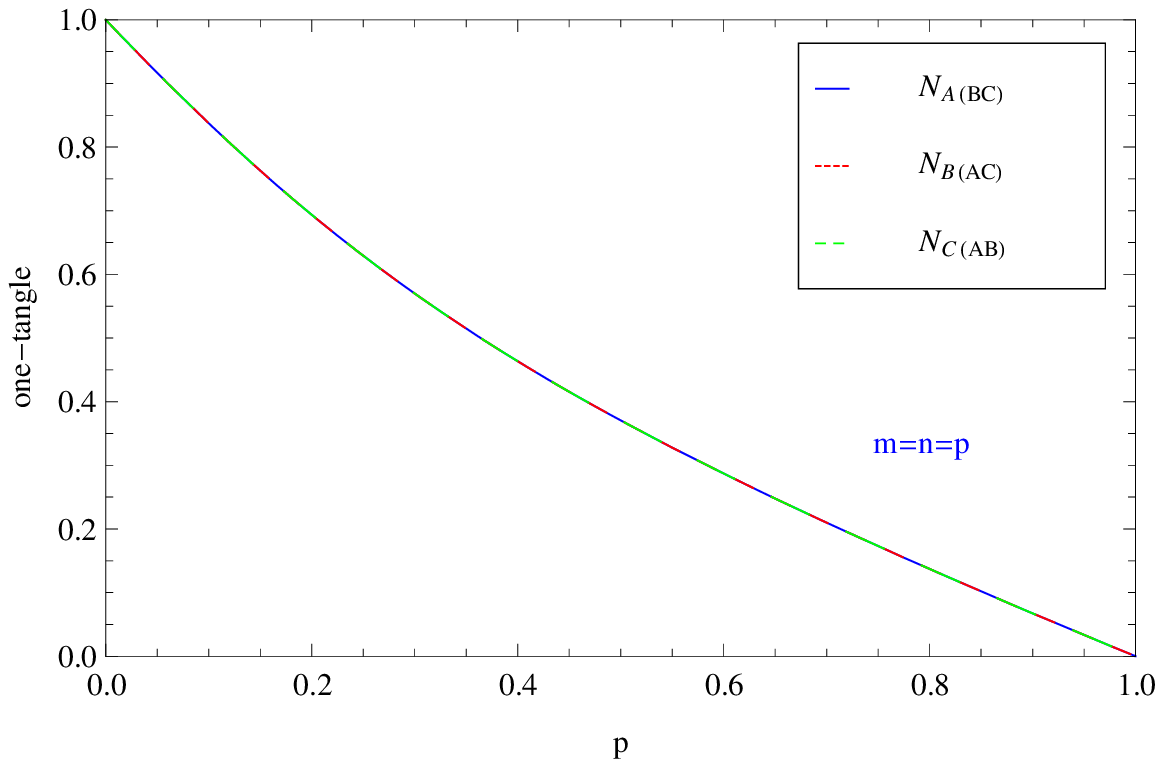}
\includegraphics[scale=0.42]{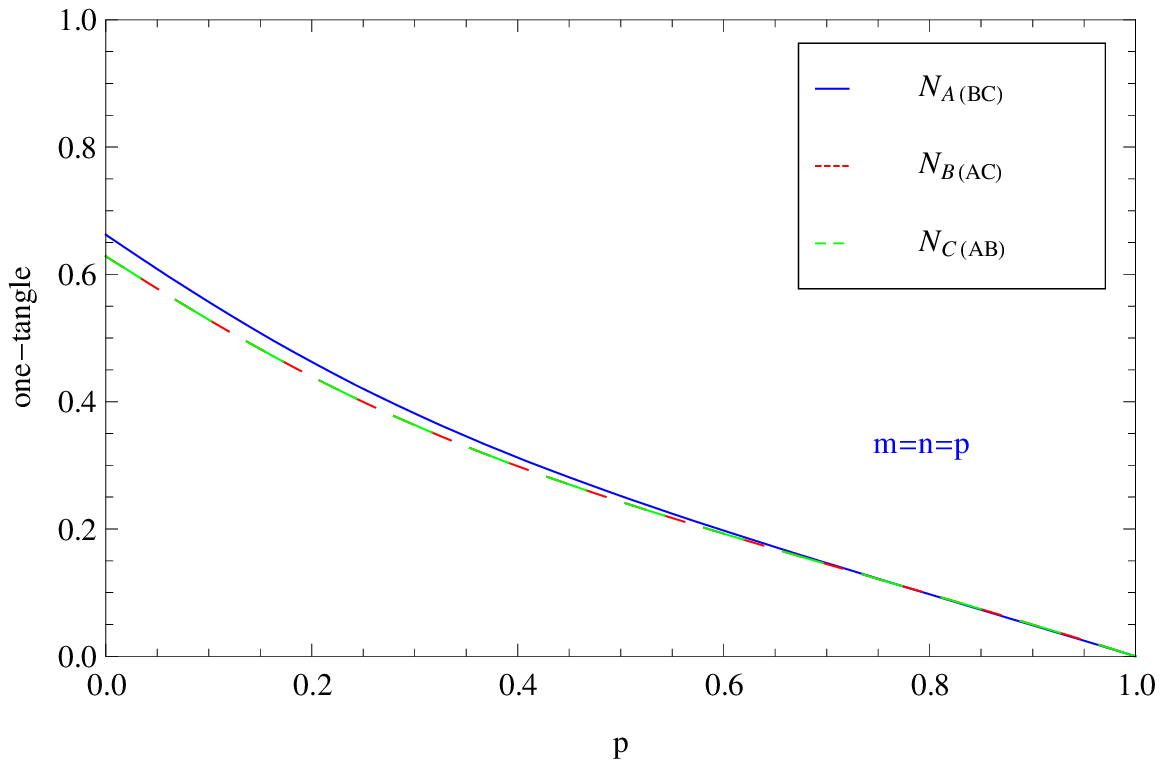}
\includegraphics[scale=0.42]{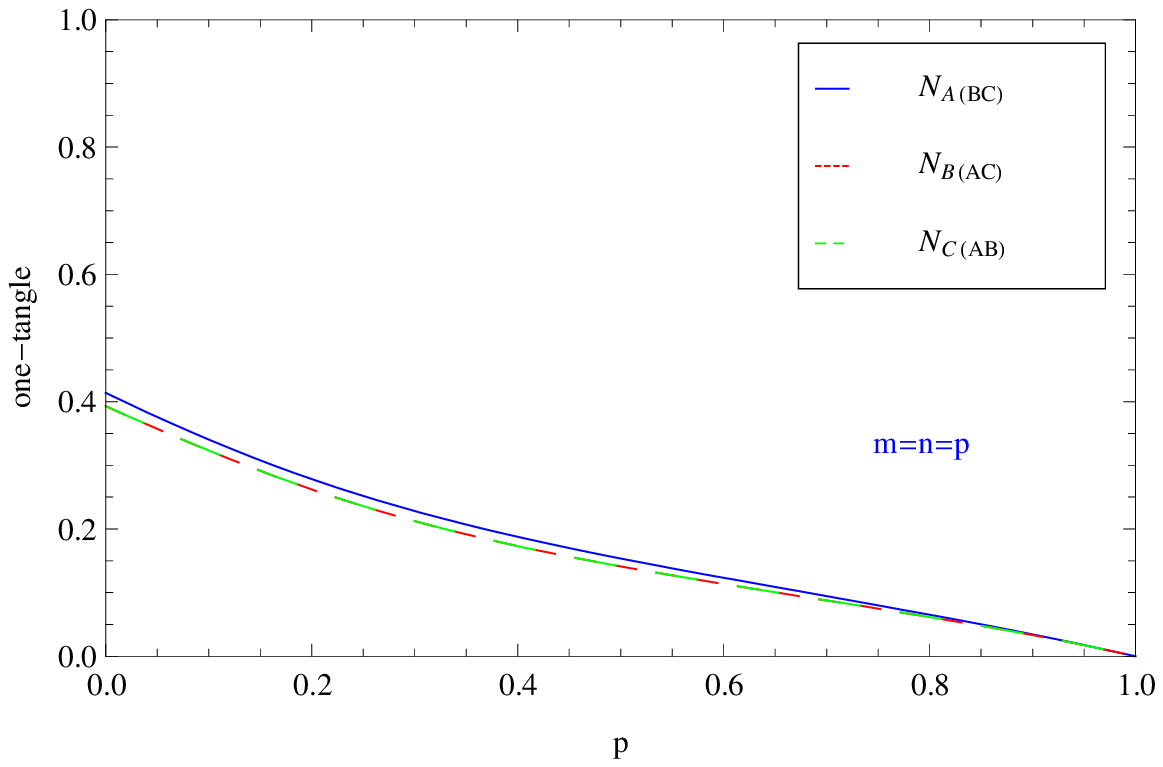}
\caption{\label{Fig.5} (Color online) The plot shows the negativity
$N_{A(B_{I}C_{I})}$ (blue line ), $N_{B_{I}(AC_{I})}$ (red line) and
$N_{C_{I}(AB_{I})}$ (green line) for amplitude damping channel. The
first (second) row presents that only inertial observer Alice
(noninertial observer Charlie) is under the environment. And the
third row for the case that Alice, Bob, and Charlie interact
with the environment. We draw them for $r=0$ (left rank), $r=\pi/6$
(middle rank), and $r=\pi/4$ (right rank). All the pictures have
consider the normalization constant $1/\sqrt{2}$. }
\end{figure}

For one-tangle, we will study three cases: $m=n=0$,  $m=p=0$ and
$m=n=p$.

We show the one-tangles with $m=n=0$ by the first row in Fig.
\ref{Fig.5}, which means only the inertial observer Alice
interacts with environment. It is very surprising to find out that
the one-tangles don't vanish even with $p=1$ which indicates that
the amplitude damping channel can't destroy the tripartite
entanglement no matter how longer it interacts with Alice. The three
subsystems still can not be distinguished at the intersect points.

The one-tangles with $m=p=0$ is found on the second row in Fig.
\ref{Fig.5}, which means only the noninertial observer Charlie
interacts with environment. If $r=0$,
$N_{B_{I}(AC_{I}})=N_{A(B_{I}C_{I})}$ as we expected. If $r\neq 0$,
the interesting result is that Bob and Charlie have the same initial
one-tangles but at last Alice and Bob have the same one-tangles,
which is different form the case of $GHZ$ state. That is to say, the
longer the time for the environment interacting with Charlie the
less the difference between Alice and Bob is, i.e., if the time is
long enough the effect of environment can wipe off the effect of
acceleration even $r=\pi/4$. The tripartite entanglement doesn't
vanish at $n=1$, either.

The situation with $m=n=p$ is shown in the third row in
Fig.\ref{Fig.5}, which means Alice, Bob, and Charlie all are under
the same environment. At $r=0$ the three subsystems can't be
distinguished as we expected. Unlike the former two cases, now we
see that the interaction with environment is strong enough to
destroy all the one-tangles when $m=n=p=1$, however, no sudden death
happens yet.

By use of Eq. (\ref{Eq.9}) we find the two-tangle between any two
subsystems of the multipartite system
\begin{eqnarray}\label{Eq.30}
N_{AB_{I}}&=&\frac{1}{6}\{-2 - 2 m - 2 p \cos{2 r} + 2 m p  \cos{2
r} +2 \sqrt{(1 - m) (1 -p) [\cos^{2}{r} + (1 - m) (1 -p)
\sin^{4}{r}]}\nonumber
\\&& +\sqrt{
   4 (1 - m) (1 -p) \cos^{2}{r} + [1 + p +
      m (3 + p) - (-1 + m) (1 + p) \cos{2 r}]^{2}} - 2 m p
   \},
\end{eqnarray}
\begin{eqnarray}\label{Eq.31}
N_{AC_{I}}&=&\frac{1}{6}\{-2 - 2 n  - 2 p \cos{2 r} +  2 n p \cos{2
r} +  2 \sqrt{(1 - n) (1 -p)[\cos^{2}{r} + (1 - n) (1 -p)
\sin^{4}{r}]} \nonumber
\\&& +
\sqrt{4 (1 - n) (1 -p) \cos^{2}{r} + [1 + p +
      n (3 + p) - (-1 + n) (1 + p) \cos{2 r}]^{2}}- 2 n p
  \},
\end{eqnarray}
\begin{eqnarray}\label{Eq.32}
N_{B_{I}C_{I}}&=&\frac{1}{12}\{ 4\cos{2 r}-8(m+n)\cos{2 r}
+ 12 m n \cos{2 r}-\cos{4 r}+(m+n)\cos{4 r}- m n \cos{4 r}
+\nonumber
\\&& 2\sqrt{\beta\gamma[4\cos^{4}{r}+\beta\gamma(-5+\cos{2 r})^{2}
\sin^{4}{r}]} -7+3 m +3 n -11 m n +\nonumber
\\&&4\sqrt{\beta\gamma\cos^{4}{r}+ [\cos^{4}{r}-\frac{1} {2}(m + n)
\cos^{2}{r}(-3+\cos{2 r})+m n\sin^{2}{r}(2 + \sin^{2}{r})]^{2}}\},
\end{eqnarray}
here $\beta=-1+m$, $\gamma=-1+n$. It is easy to find that
$N_{AB_{I}}$($N_{AC_{I}}$, and $N_{B_{I}C_{I}}$) is independent on
$n$ ($m$, and $p$), i.e., a subsystem interacts with environment
wouldn't affect the two-tangle between the other two subsystems. And
all the initial two-tangles don't equal to zero which is different
from the case of $GHZ$ state.

\begin{figure}[ht]
\includegraphics[scale=0.43]{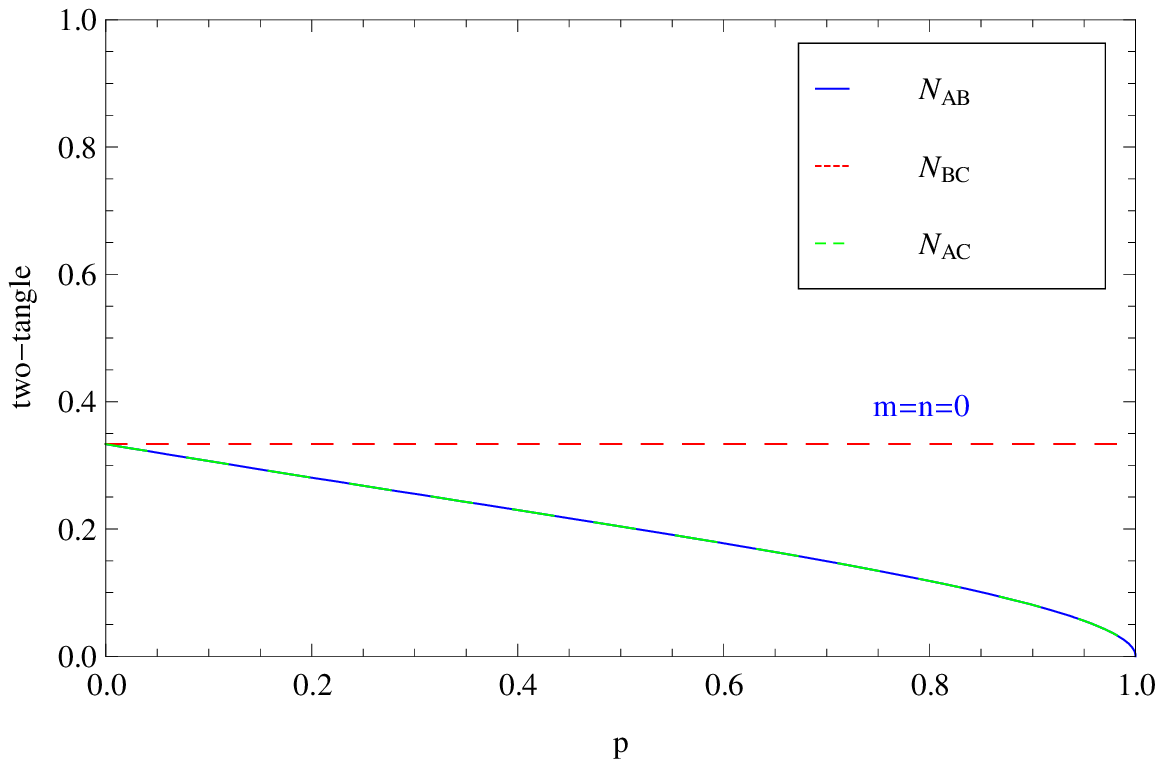}
\includegraphics[scale=0.42]{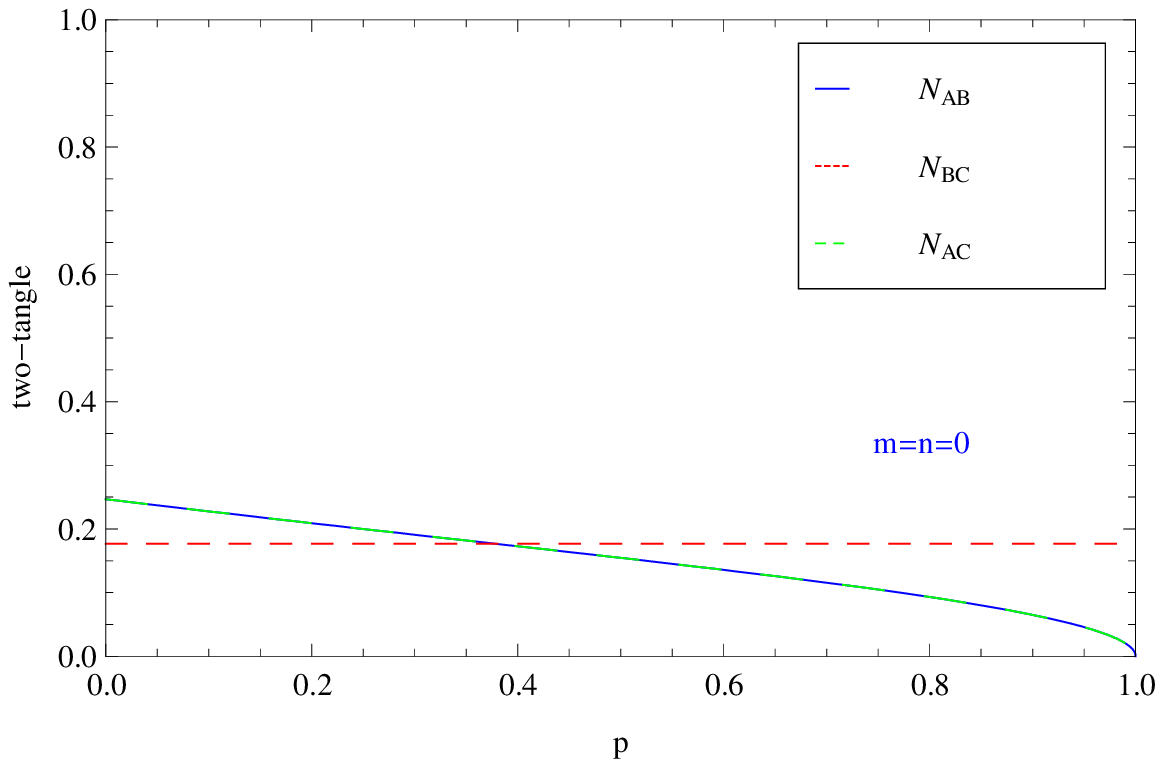}
\includegraphics[scale=0.42]{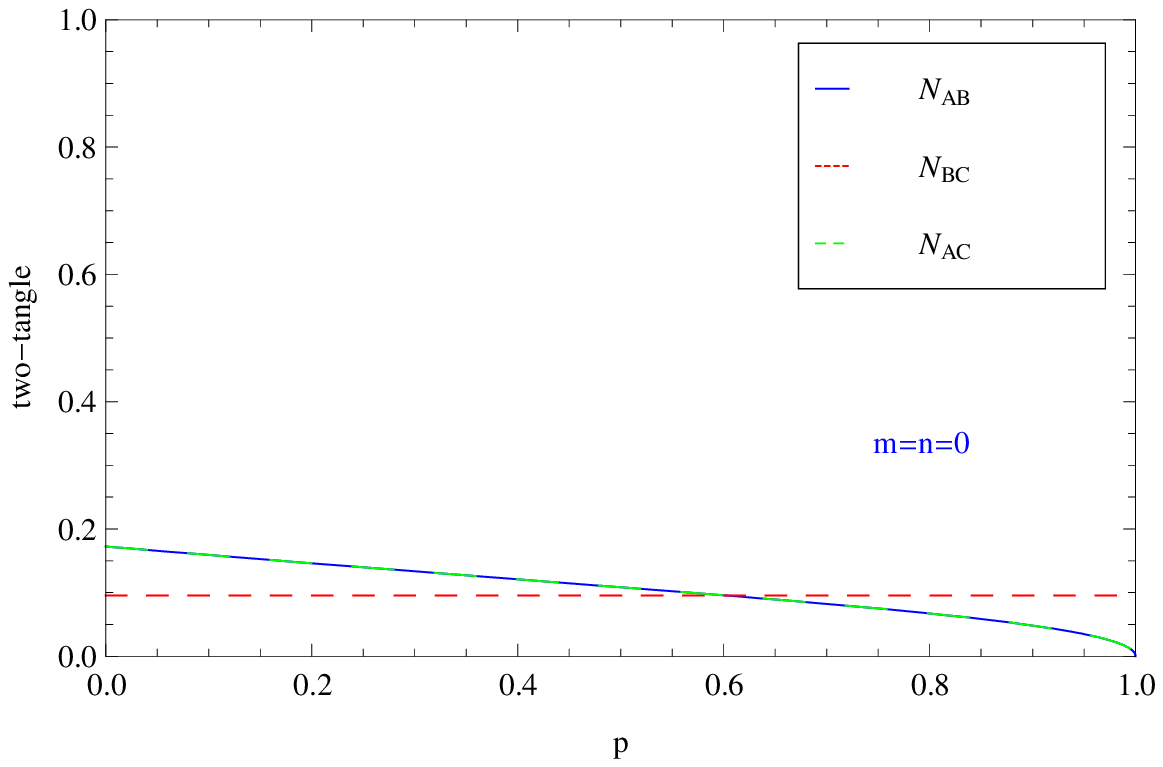}\\
\includegraphics[scale=0.43]{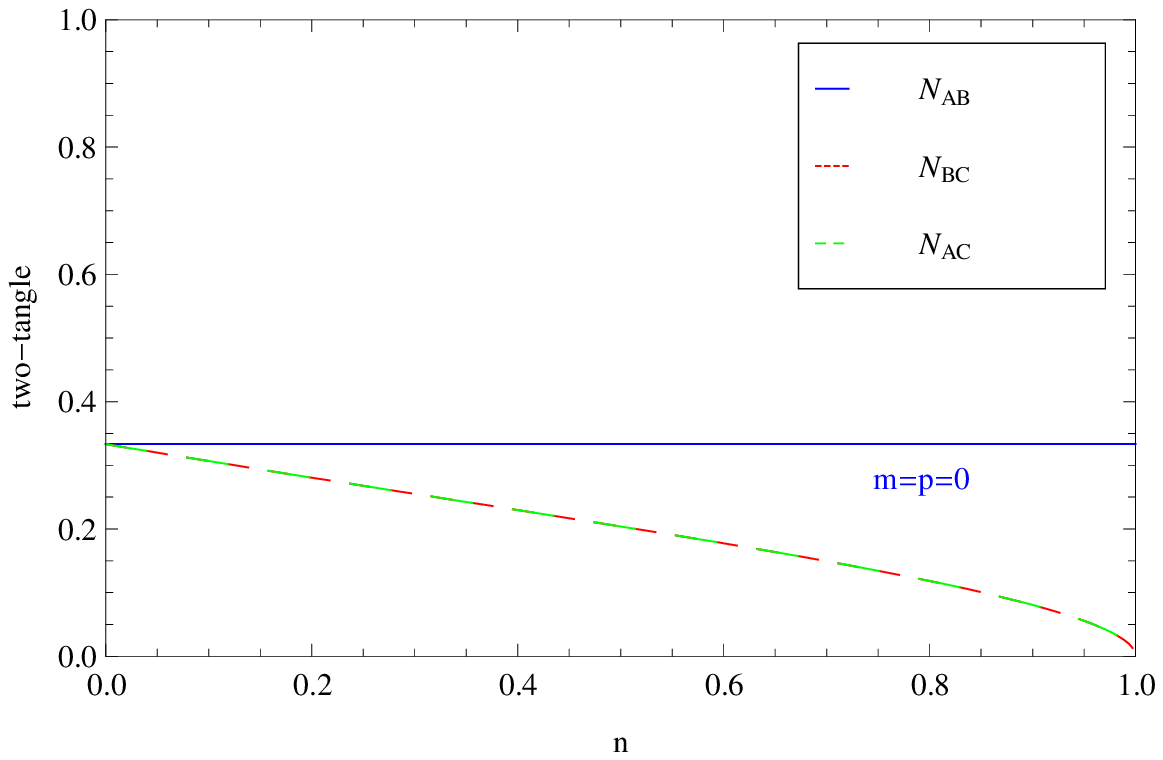}
\includegraphics[scale=0.42]{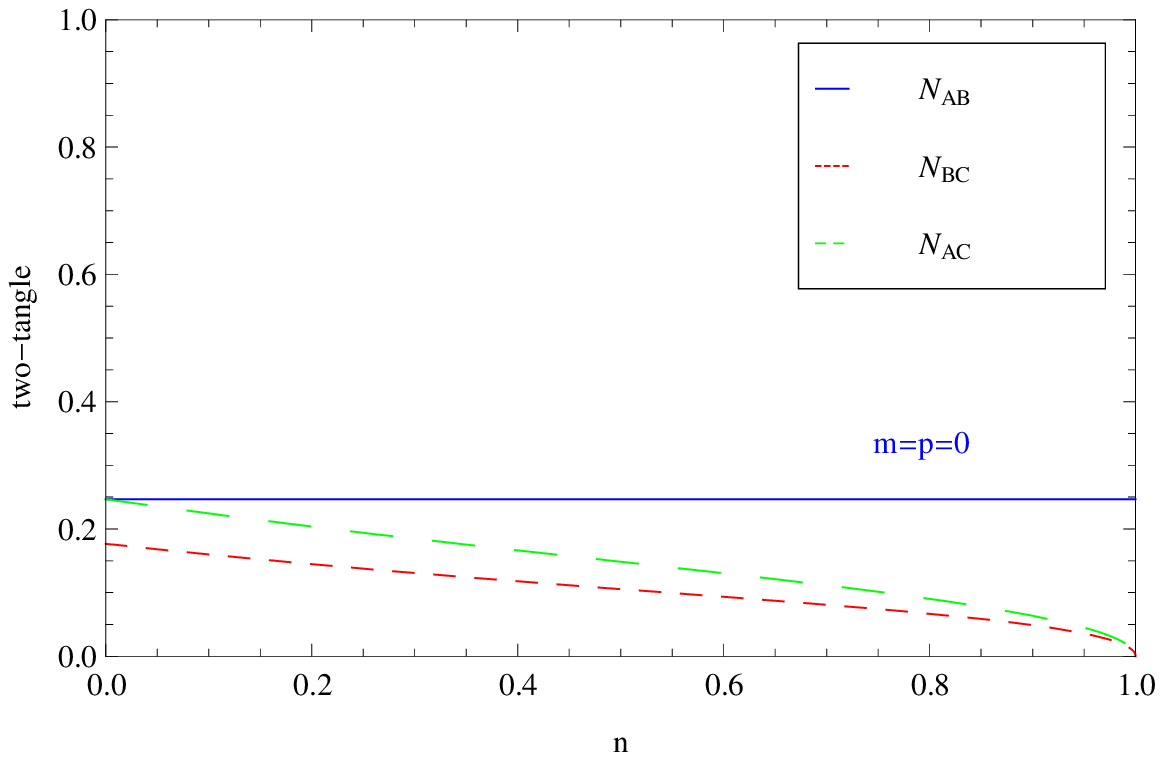}
\includegraphics[scale=0.42]{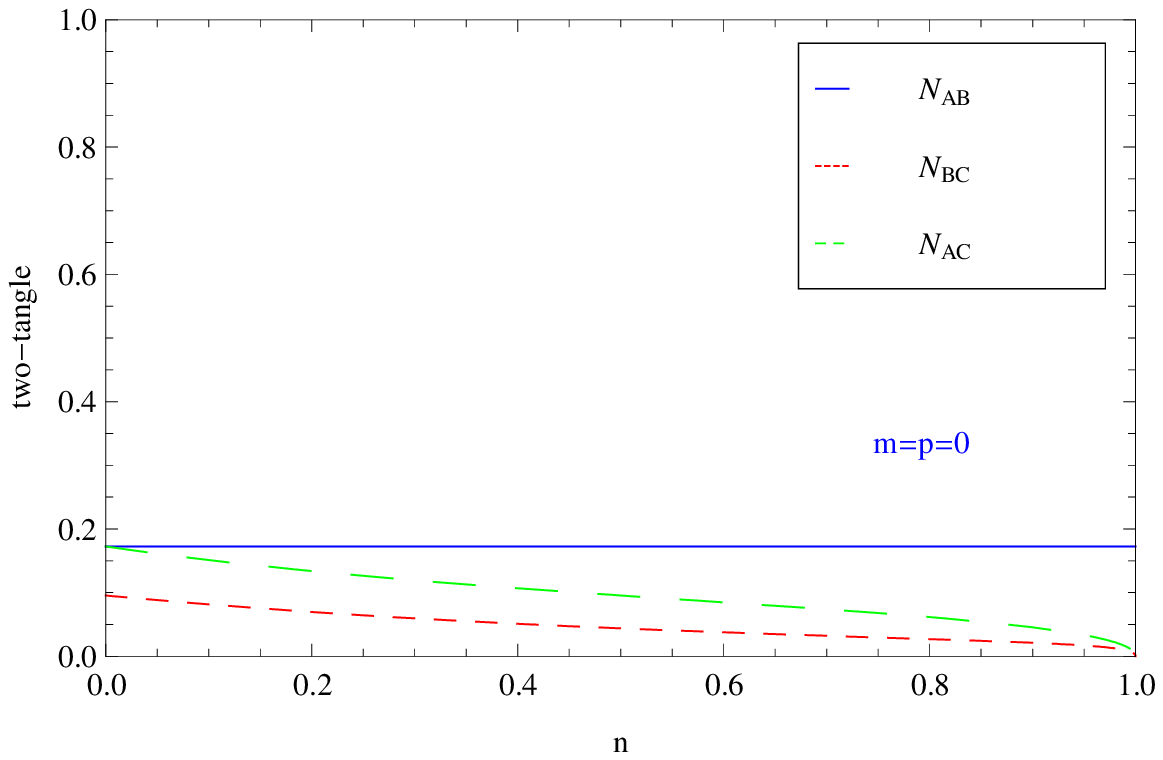}\\
\includegraphics[scale=0.43]{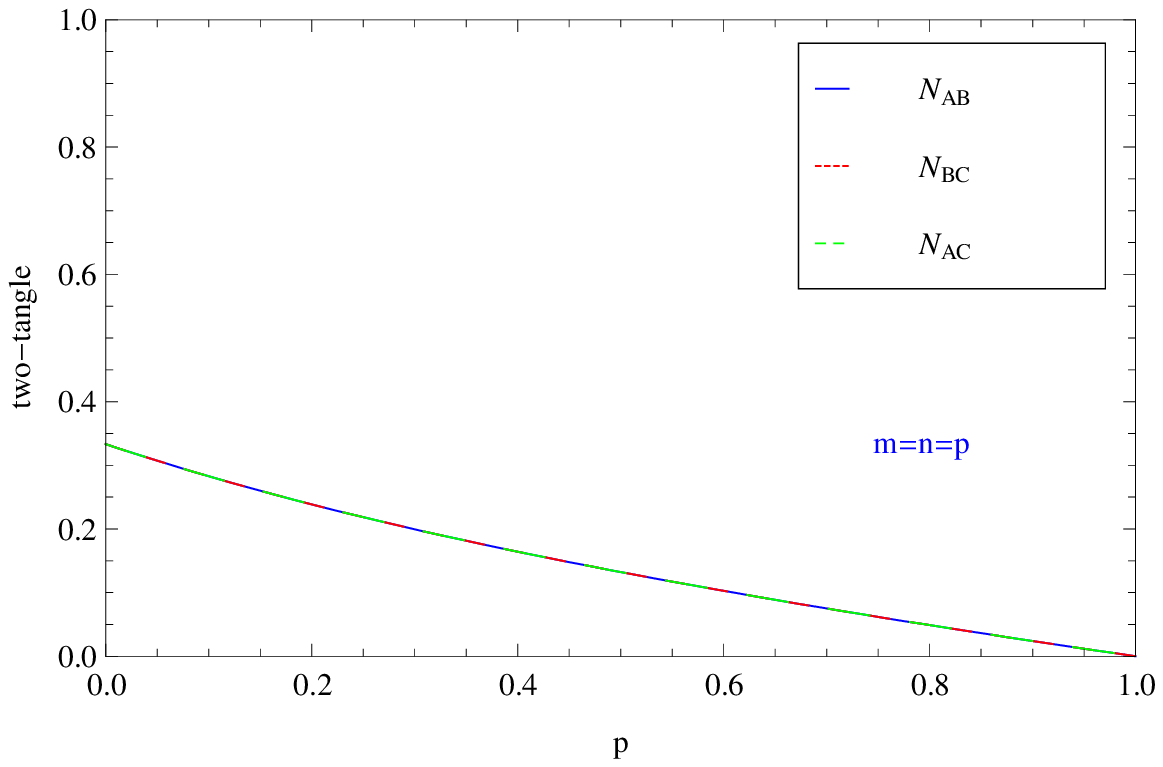}
\includegraphics[scale=0.42]{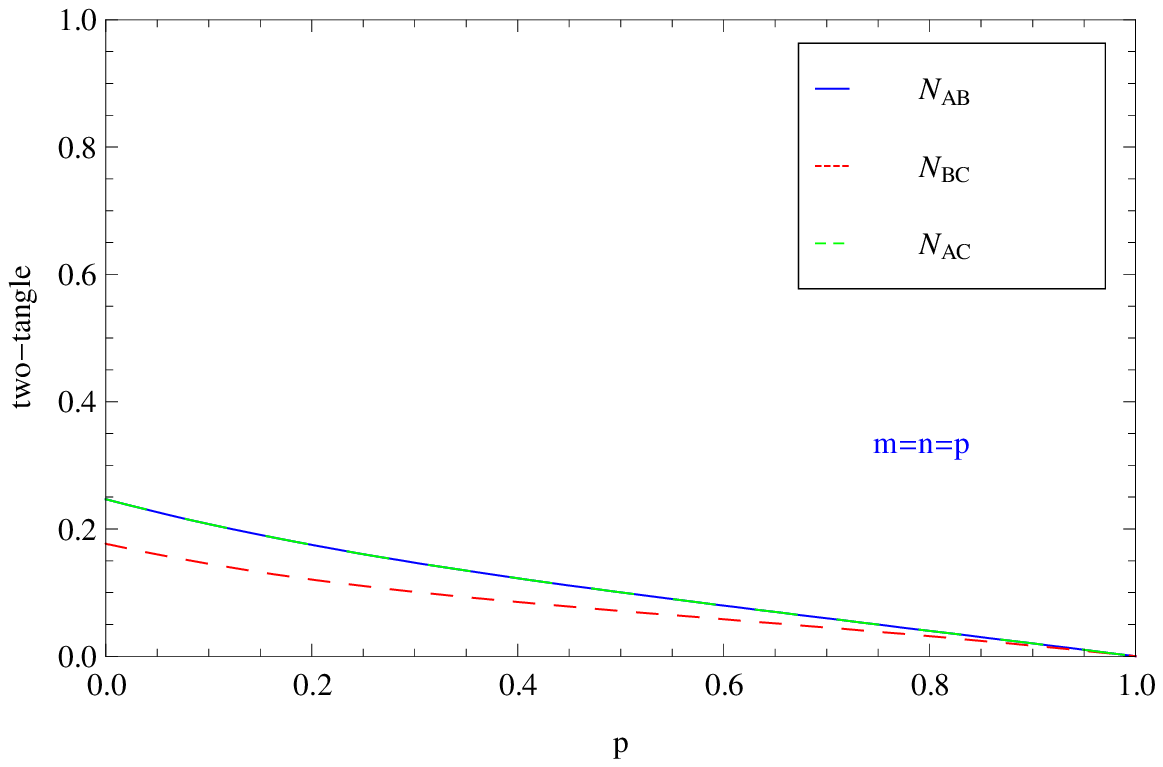}
\includegraphics[scale=0.42]{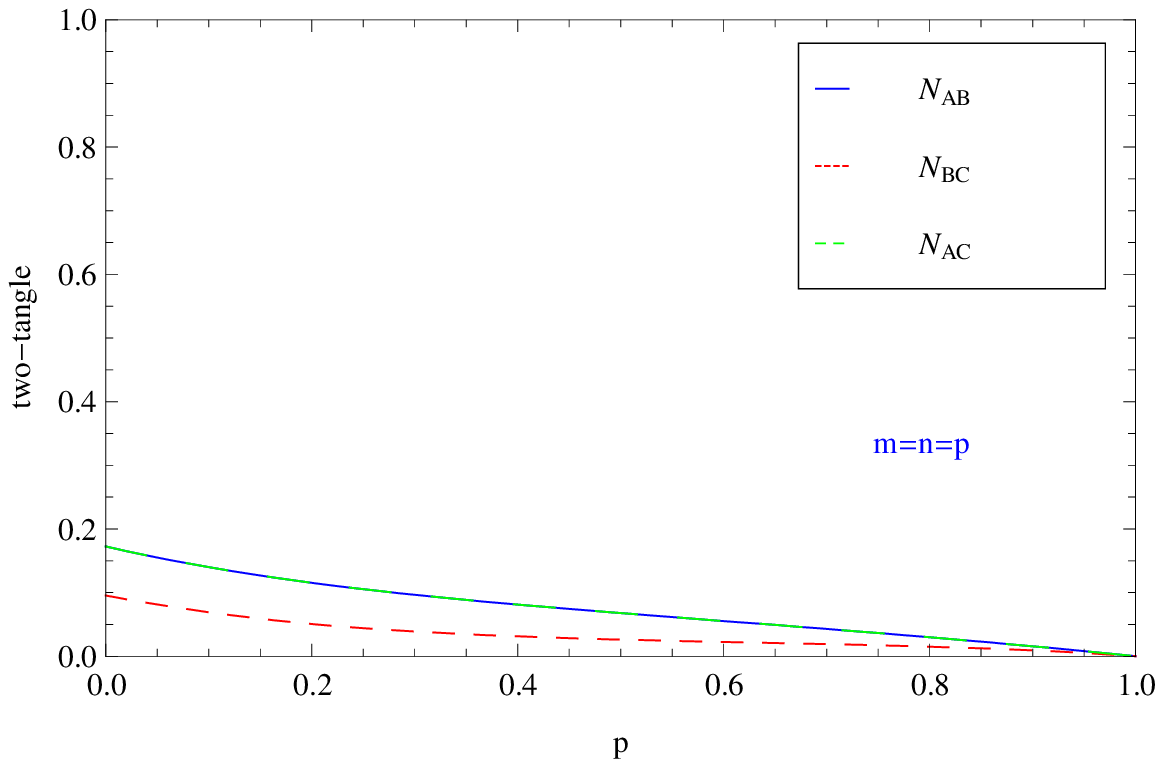}
\caption{\label{Fig.11}(Color online) The plot shows the negativity
$N_{AB_{I}}$ (blue line), $N_{BC_{I}}$ (red line), and $N_{AC_{I}}$
(green line) for amplitude damping channel. The first (second) row
presents that only inertial observer Alice (noninertial observer
Charlie) is under the environment. And the third row for the case
that Alice, Bob, and Charlie all interact with the environment. We
draw them for $r=0$ (left rank), $r=\pi/6$ (middle rank), and
$r=\pi/4$ (right rank). All the pictures have considered the
normalization constant $1/\sqrt{2}$. }
\end{figure}

For two-tangle, we will also study three cases: $m=n=0$,  $m=p=0$
and $m=n=p$.

We give the results with $m=n=0$ on the first row in Fig.
\ref{Fig.11}, which indicates only Alice is under the environment.
Note that $N_{B_{I}C_{I}}$ is a constant and
$N_{AB_{I}}=N_{AC_{I}}$, but unlike the one-tangles they both vanish
in the limit of $p=1$, i.e., the environment can destroy the
two-tangles completely just like the case of two-qubits. But the
difference is that no sudden death happens yet.

We show the two-tangles with $m=p=0$ at the second row in  Fig.
\ref{Fig.11}, which indicates only Charlie is under the environment.
It is found that $N_{AC_{I}}=N_{AB_{I}}$ when $n=0$ (i.e., without
considering the environment). If $n>0$, the environment will destroy
the symmetry between Bob and Charlie as we expected, and the
acceleration will destroy the symmetry between Alice and Bob. The
two-tangles $N_{AC_{I}}$ and $N_{B_{I}C_{I}}$ disappear completely
when $n=1$.

The two-tangles with $m=n=p$ is shown in the last row in Fig.
\ref{Fig.11}, which means all the subsystems are under the
environment. Now all of the two-tangles are destroyed since all of
subsystems are under the environment. Note that
$N_{AB_{I}}=N_{AC_{I}}$ as we expected.

\begin{figure}[ht]
\includegraphics[scale=0.43]{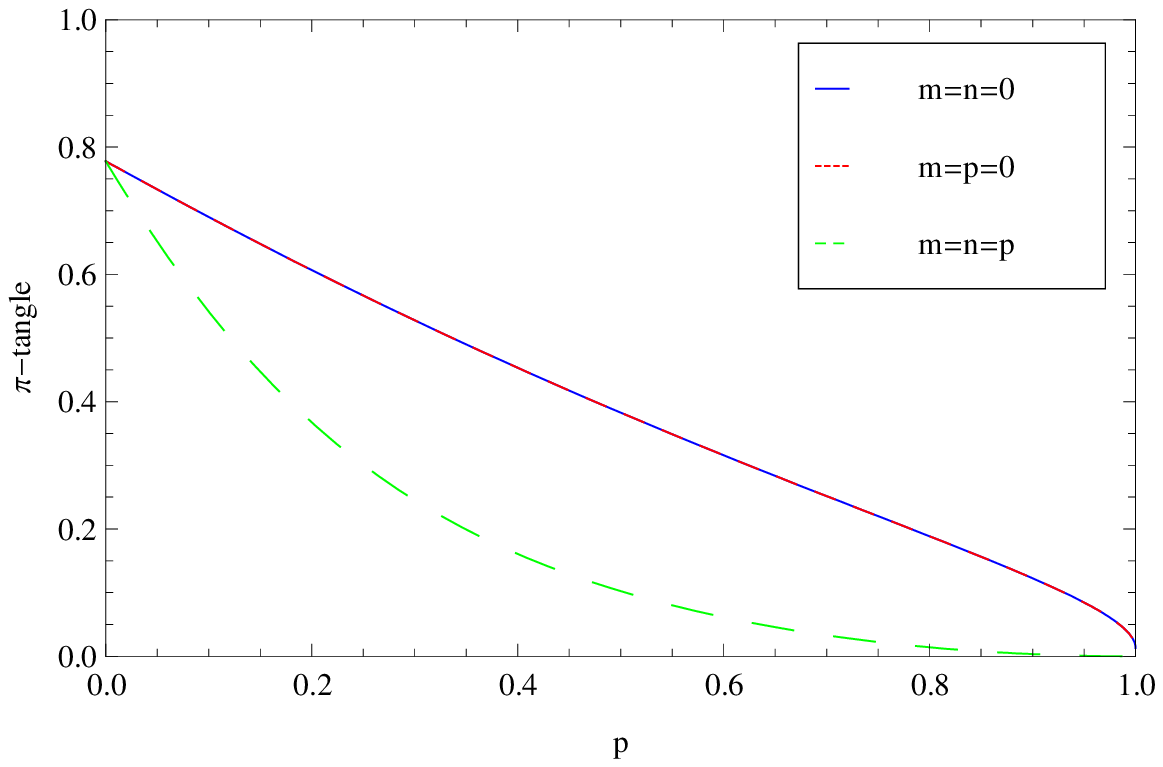}
\includegraphics[scale=0.42]{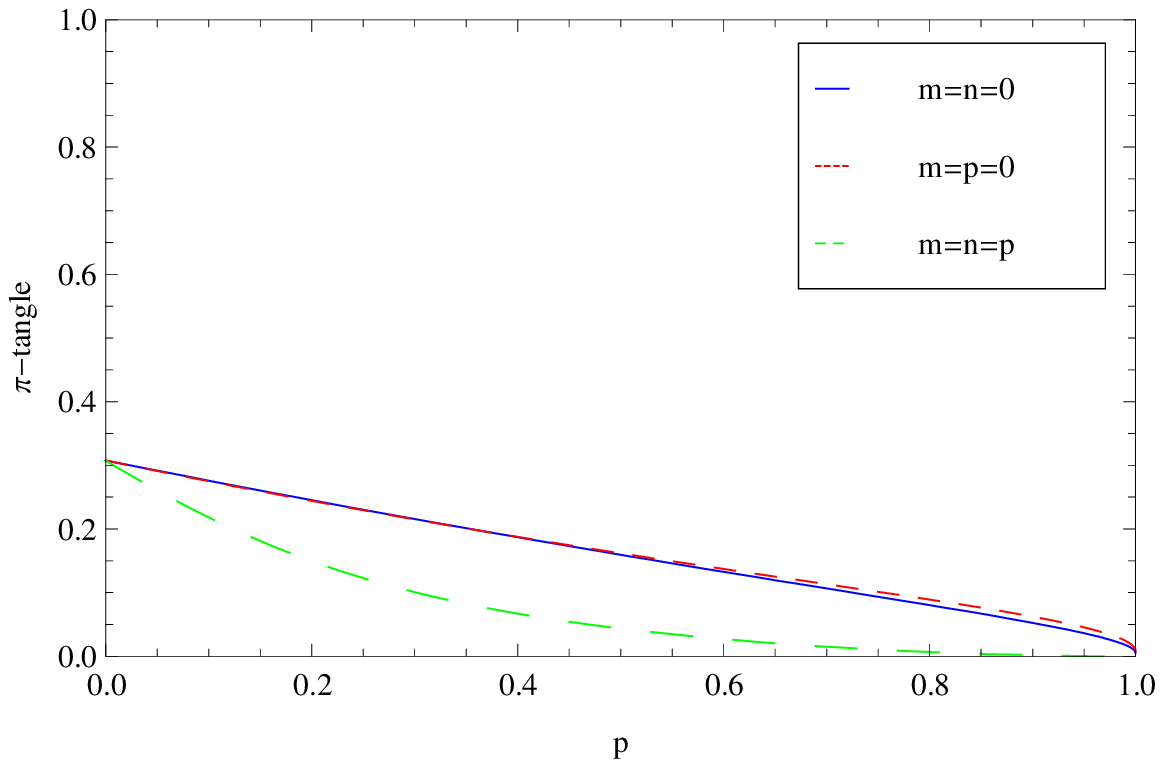}
\includegraphics[scale=0.42]{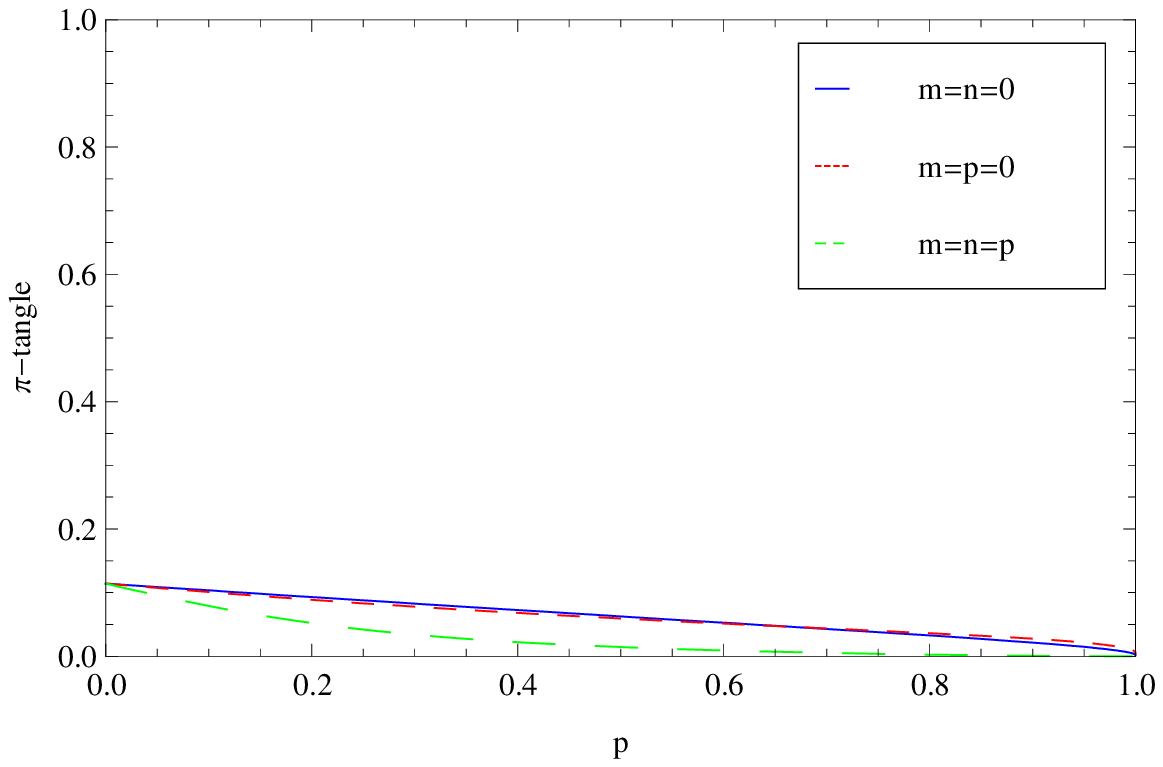}
\caption{\label{Fig.14}(Color online) The $\pi$-tangle which
considers the environment. Blue (Red) line plots the case of only
Alice (Charlie) interacting with the environment and green one
corresponds to the case of all of them interacting with the
environment. We also show three cases for $r=0$ (left), $r=\pi/6$
(middle), and $r=\pi/4$ (right). All the pictures have considered
the normalization constant $1/\sqrt{2}$.}
\end{figure}
At last, we compute the $\pi$-tangle by use of Eqs. (\ref{Eq.11})
and (\ref{Eq.12}) meanwhile consider the normalization constant
$1/\sqrt{2}$. The result is found in Fig. \ref{Fig.14}, just like
before, the effect of environment is much more stronger than the
effect of acceleration. For $W$ state, the initial $\pi$-tangle is
smaller than that for $GHZ$ state because there exist two-tangles.

\subsection{Depolarizing noise}

\begin{figure}[ht]
\includegraphics[scale=0.42]{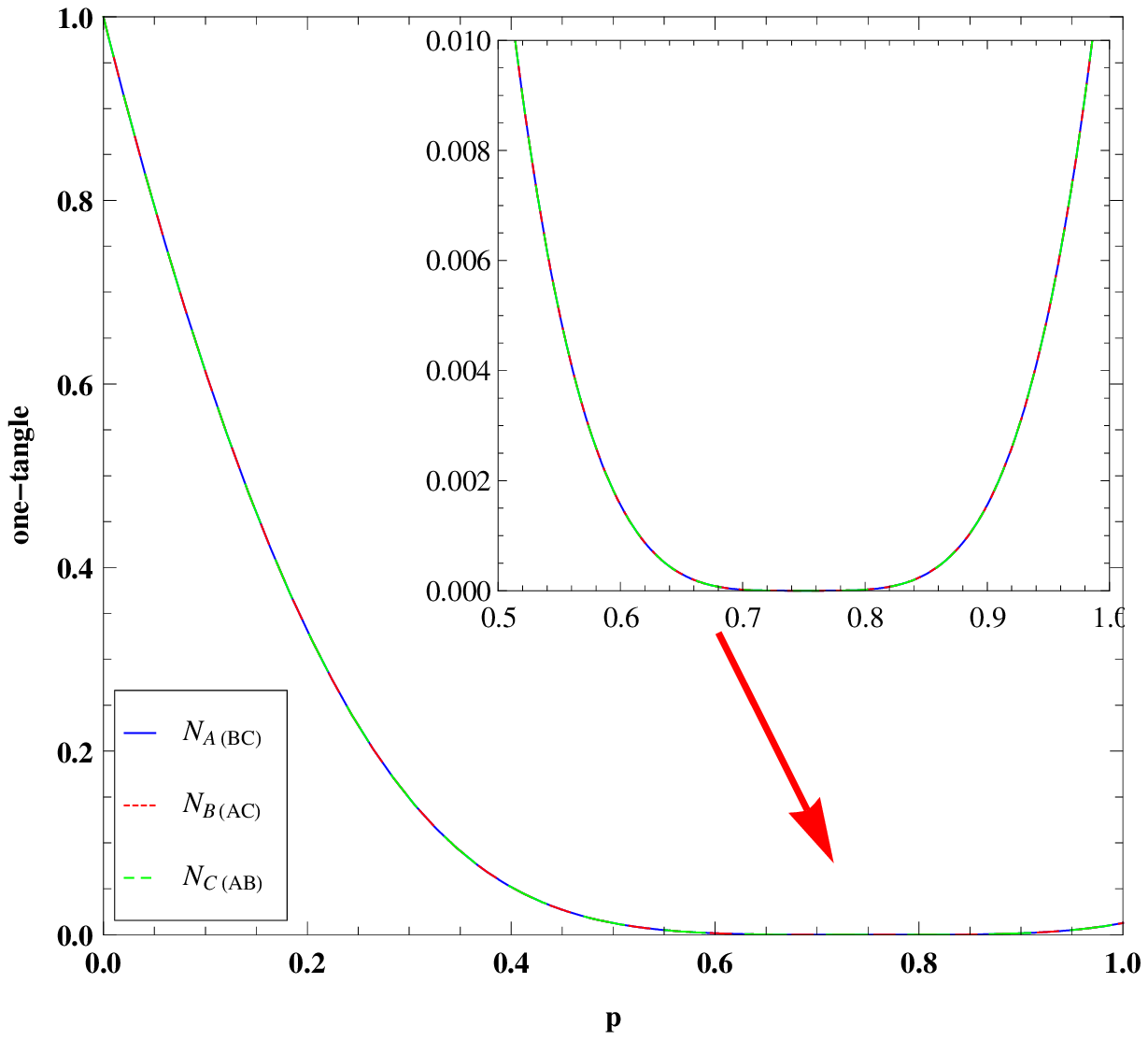}
\includegraphics[scale=0.42]{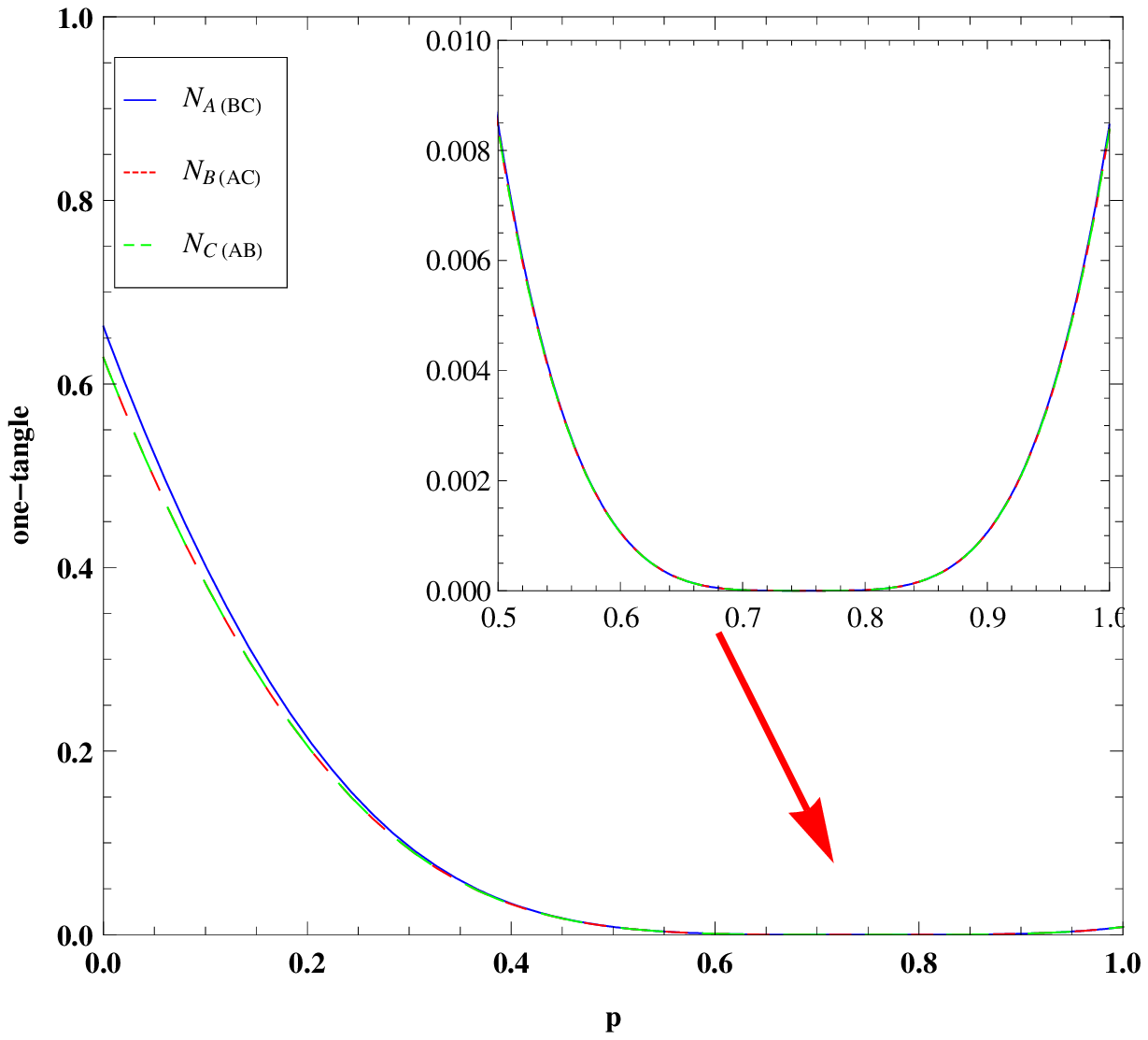}
\includegraphics[scale=0.42]{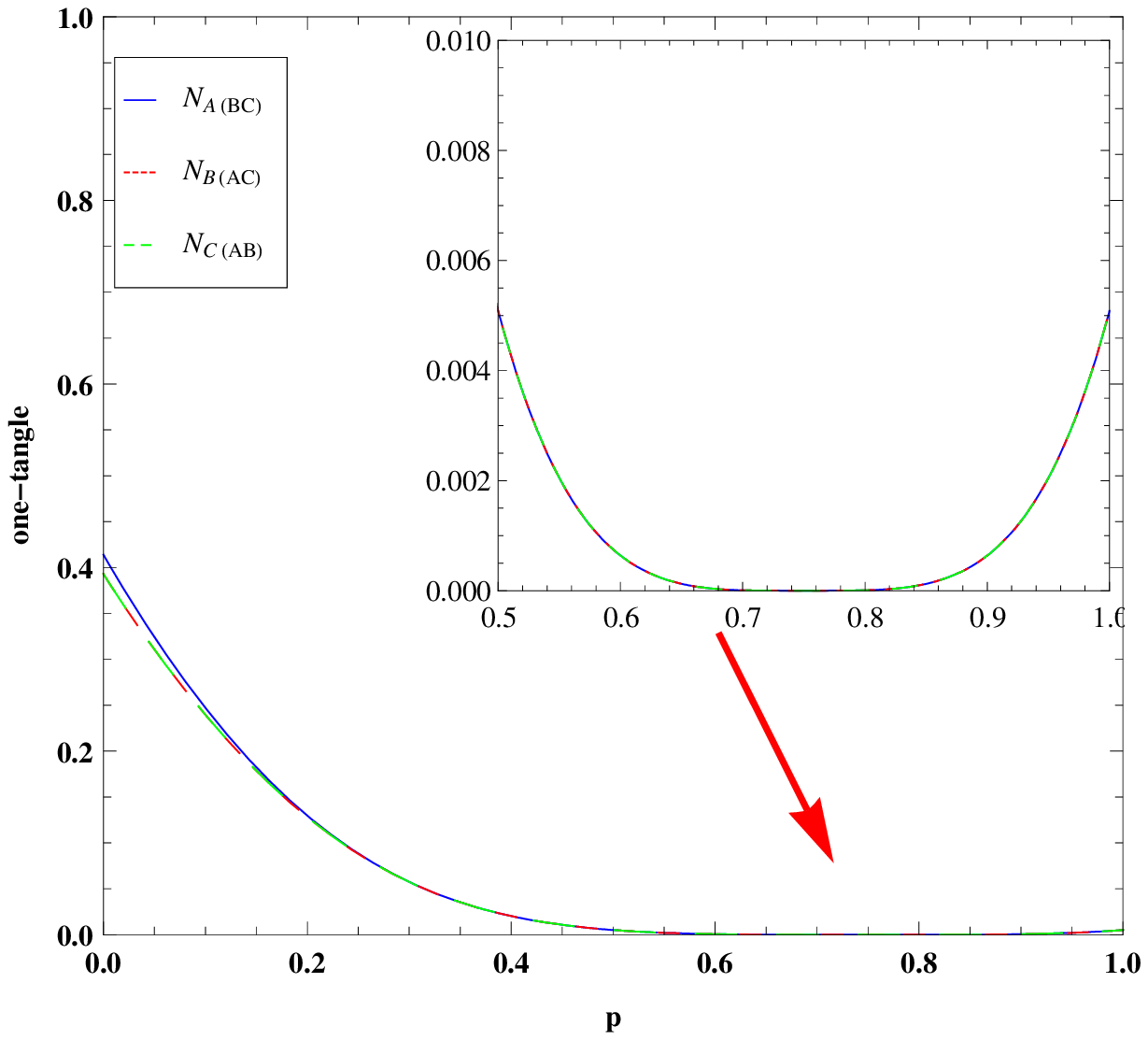}
\caption{\label{Fig.15}(Color online) The negativity
$N_{A(B_{I}C_{I})}$ (blue line ), $N_{B_{I}(AC_{I}})$ (red line),
and $N_{C_{I}(AB_{I})}$ (green line) when Alice, Bob, and Charlie
all are in depolarizing noise. We show three cases for $r=0$ (left),
$r=\pi/6$ (middle), and $r=\pi/4$ (right). The rebound process is
plotted in the magnifying pictures. All the pictures have considered
the normalization constant $1/\sqrt{2}$.}
\end{figure}

Repeating the foregoing steps and using Eqs. (\ref{Eq.3}),
(\ref{Eq.8}) and (\ref{Eq.10}), we give the results in Fig.
\ref{Fig.15} for case that all the subsystems in depolarizing noise.
We see that many former characteristics still remain under this
environment, too. And the rebound process is much more stronger than
before when $p>0.75$ and we can hardly ignore it any more.
Similarly, the acceleration can resist the rebound process but can't
destroy it. And we predict that this process would be more stronger
in a higher dimensionality.

\begin{figure}[ht]
\includegraphics[scale=0.42]{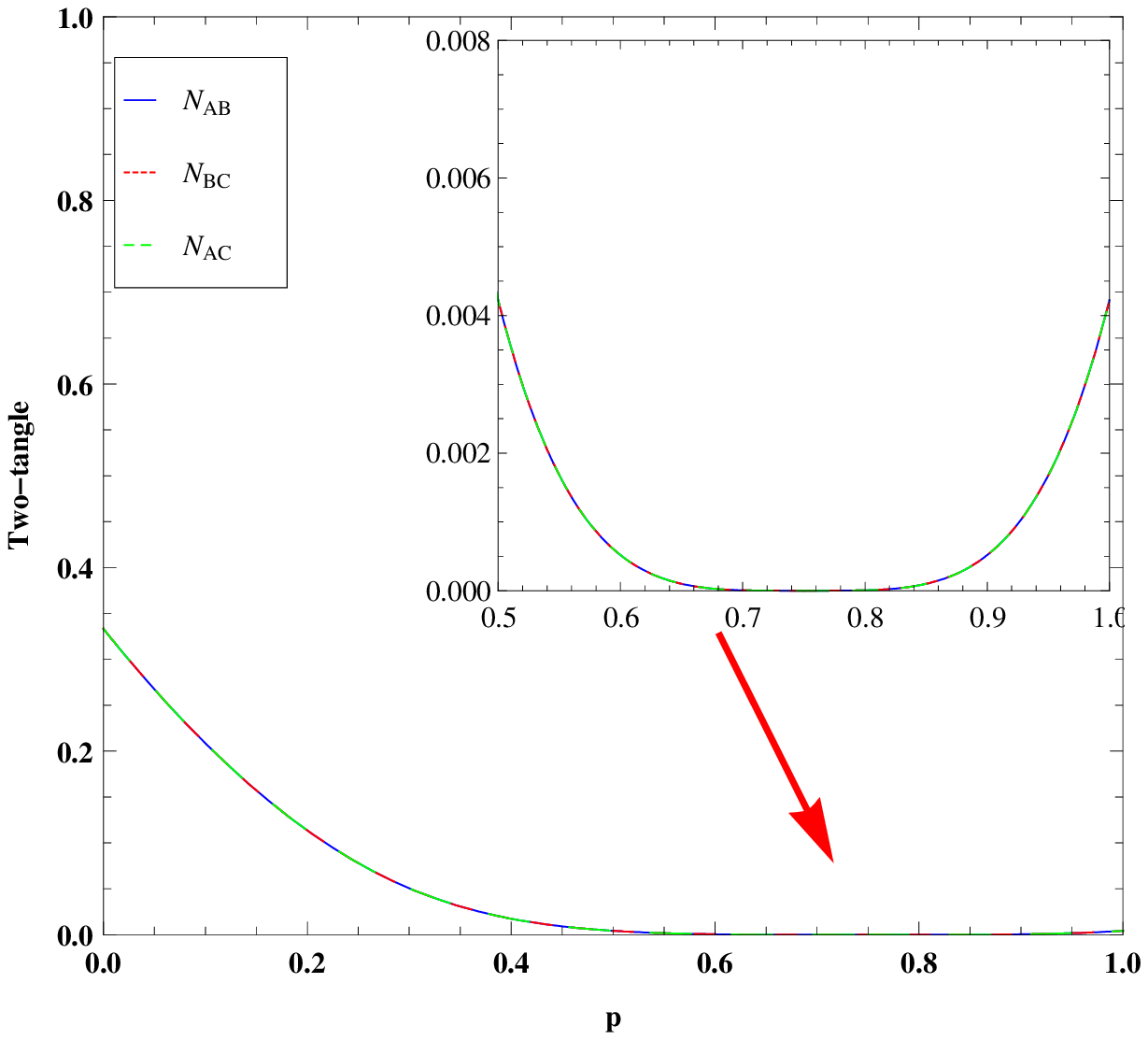}
\includegraphics[scale=0.42]{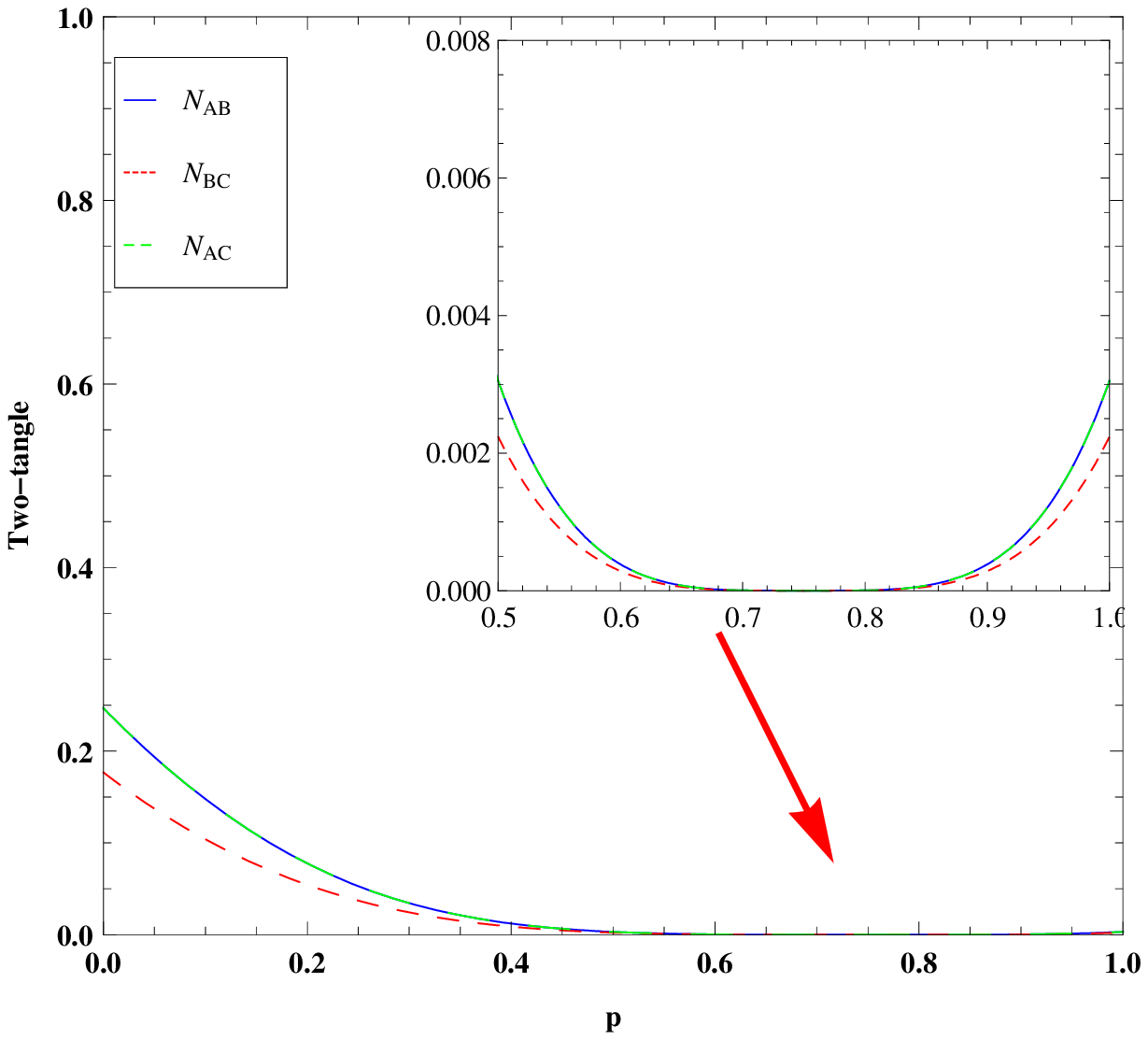}
\includegraphics[scale=0.42]{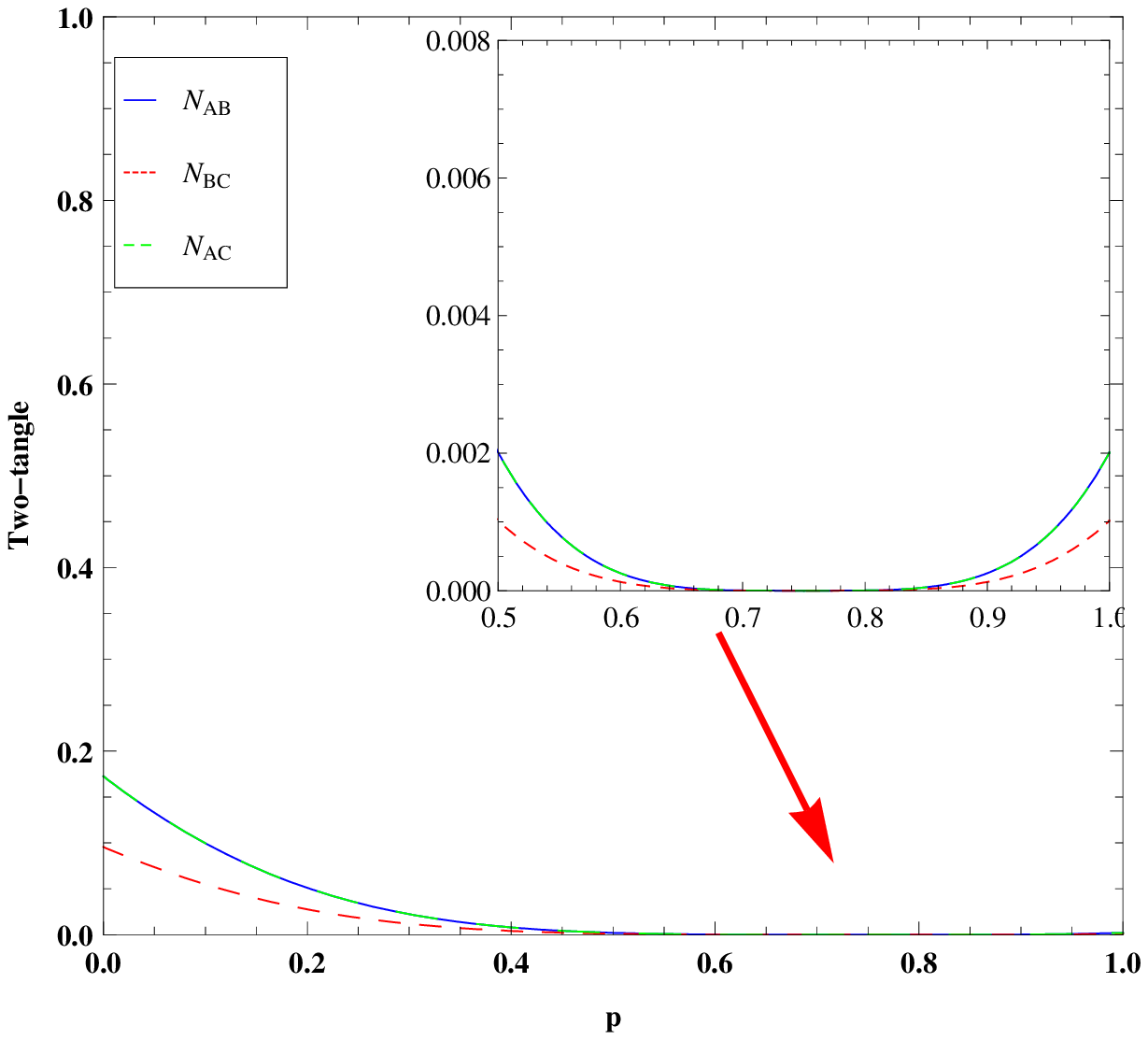}
\caption{\label{Fig.16}(Color online) The negativity  $N_{AB_{I}}$
(blue line ), $N_{B_{I}C_{I}}$ (red line), and $N_{AC_{I}}$ (green
line) when all the subsystems are under the environment. We give the
cases for $r=0$ (left), $r=\pi/6$ (middle), and $r=\pi/4$ (right).
The rebound process is plotted in the magnifying pictures. All the
pictures have considered the normalization constant $1/\sqrt{2}$.}
\end{figure}

\begin{figure}[ht]
\includegraphics[scale=0.5]{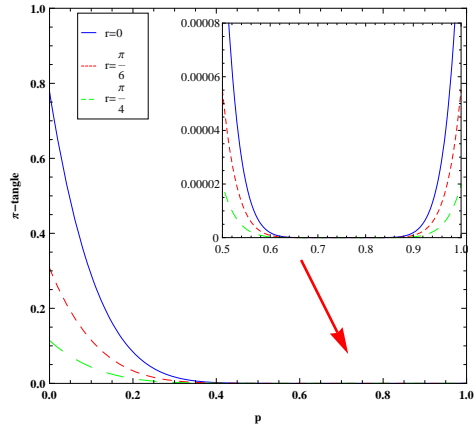}
\caption{\label{Fig.17}(Color online) The $\pi$-tangle
$\pi_{AB_{I}C_{I}}$ when Alice, Bob, and Charlie all are in
depolarizing noise. We show three cases for $r=0$ (dotted line ),
$r=\pi/6$ (dash line), and $r=\pi/4$ (solid line). All the pictures
have considered the normalization constant $1/\sqrt{2}$.}
\end{figure}

In addition, using Eq. (\ref{Eq.9}) we give $N_{AB_{I}}$,
$N_{AC_{I}}$ and $N_{B_{I}C_{I}}$ in Fig. \ref{Fig.16}. What's
surprising is that in the tripartite system the two-tangles also
have a rebound process in depolarizing noise when $p>0.75$. The
$\pi$-tangle and its rebound process are plotted in Fig.
\ref{Fig.17}. Now we can say that unlike the case of two-qubits
there is no sudden death even all the subsystems are in the
depolarizing noise. But an entanglement rebound process appears.

We note again that the CKW inequality \cite{23},
$N_{AB}^{2}+N_{AC}^{2}\leq N_{A(BC)}^{2}$, is still saturated for
this state, which means the effect of environment and the
noninertial frames don't destroy this inequality for $W$ initial
state, either.

\section{SUMMARY}
The tripartite entanglement of a 3-qubit fermionic  system under the
amplitude damping channel and in depolarizing noise when two
subsystems are accelerated for the $GHZ$ and $W$ initial states is
investigated. It is shown that all the one-tangles and $\pi$-tangles
decrease more quickly when subsystems are under environment.
However, unlike the case of 2-qubit system in which sudden death can
be taken place easily, here a surprising result is that no sudden
death happens for any acceleration even all the subsystems are under
the environment. We can't distinguish all the subsystems when
$p=\cos{2r}\sin^{2}{r}$ if only Alice is under the amplitude
environment for the $GHZ$ state. All the entanglement decreases more
quickly in depolarizing noise than that in amplitude damping
environment. It is found that no bipartite entanglement generates
either in the accelerated subsystem or under the environment for the
$GHZ$ state, i.e., all the entanglement is in form of tripartite
entanglement in this case. But bipartite entanglement exists for the
$W$ state. Both the effect of acceleration and environment can
destroy the symmetry between the subsystems. Thus we can perform
such quantum information tasks to distinguish the
accelerated-observers when some observers are accelerating by using
the effect of environment or distinguish some observers in the same
environment with the effect of acceleration. We furthermore give a
conclusion that the more strong the subsystem interacts with the
environment is, the faster the entanglement decays. And the effect
of environment is so strong that we can nearly ignore the effect of
acceleration if the time is long enough. In depolarizing noise
environment and for both the $GHZ$ and $W$ initial states, the
entanglement will decay to zero at $p=0.75$ and then a rebound
process takes place when $p>0.75$, which means that all the
tripartite entanglement transfers to environment at $p=0.75$ and
then part of it transfers from environment back to the system when
$p>0.75$.  The CKW inequality $N_{AB}^{2}+N_{AC}^{2}\leq
N_{A(BC)}^{2}$ is saturated for any case in this paper, which means
the effects of environment and acceleration don't destroy this
inequality.

\vspace*{2.0cm} {\it Acknowledgments:} This work was supported by
the  National Natural Science Foundation of China under Grant No
10875040;  a key project of the National Natural Science Foundation
of China under Grant No 10935013;  the National Basic Research of
China under Grant No. 2010CB833004,  PCSIRT under Grant No. IRT0964,
and the Construct Program  of the National Key Discipline.


\vspace*{0.5cm}



\begin{thebibliography}{99}

\bibitem{1} M. A. Nielsen and I. L. Chuang, Quantum  Computation and Quantum Information (Cambridge University Press, Cambridge, England, 2000).

\bibitem{2} C. H. Bennett, G. Brassard, C. Cr\'{e}peau,  R. Jozsa, A. Peres, and W. K. Wootters, Phys. Rev. Lett. {\bf 70},1895 (1993).

\bibitem{3} C. H. Bennett and S. J. Wiesner, Phys. Rev. Lett. {\bf 69}, 2881 (1992).

\bibitem{4} A. K. Ekert, Phys. Rev. Lett. {\bf 67}, 661 (1991).

\bibitem{5} M. Curty, M. Lewenstein, and N. Lutkenhaus, Phys. Rev. Lett. {\bf 92}, 217903 (2004).

\bibitem{6} T. Yu and J. H. Eberly, Phys. Rev. Lett. {\bf 97}, 140403 (2006).

\bibitem{7} T. Yu and J. H. Eberly, Phys. Rev. Lett. {\bf 93}, 140404 (2006).

\bibitem{8} K. Zyczkowski, P. Horodecki, M. Horodecki, and R. Horodecki, Phys. Rev. A {\bf 65}, 012101 (2001).

\bibitem{9} P. J. Dodd and J. J. Halliwell, Phys. Rev. A {\bf 69}, 052105 (2004).

\bibitem{10} C. E. L\'{o}pez, G. Romero, F. Lastra, E. Solano, and J. C. Retamal, Phys. Rev. Lett. {\bf 101}, 080503 (2008).

\bibitem{11} D. C. M. Ostapchuk and R. B. Mann, Phys. Rev. A {\bf 79}, 042333 (2009).

\bibitem{12} J. Wang, J. Deng, and J. Jing, Phys. Rev. A {\bf 81}, 052120 (2010).

\bibitem{13} R. B. Mann and V. M. Villalba, Phys. Rev. A {\bf 80}, 022305 (2009).

\bibitem{14} J. Wang, Q. Pan, S. Chen, and J. Jing, Phys. Lett. B {\bf 677}, 186 (2009).

\bibitem{15} J. Le\'{o}n and E. Martn-Martnez, Phys. Rev. A {\bf 80}, 012314 (2009).

\bibitem{16} M.-R. Hwang, D. Park, and E. Jung, Phys. Rev. A {\bf 83}, 012111 (2010).

\bibitem{17} J. Wang and J. Jing, Phys. Rev. A {\bf 83}, 022314 (2011).

\bibitem{18} M. AsPachs, G. Adesso, and I. Fuentes, Phys. Rev. Lett. {\bf 105}, 151301 (2010).

\bibitem{19} E. Mart\'{\i}n-Mart\'{\i}nez, L. J. Garay, and J. Le\'{o}n, Phys. Rev. D {\bf 82}, 064006 (2010); Phys. Rev. D {\bf 82}, 064028 (2010).

\bibitem{20} D. E. Bruschi, J. Louko, E. Martn-Martnez, A. Dragan, and I. Fuentes, Phys. Rev. A {\bf 82}, 042332 (2010).

\bibitem{21} A.  Salles,  F.  de Melo1,  M.  P.  Almeida1,  M.  Hor-Meyll,  S.  P.  Walborn,
P.  H.  SoutoRibeiro,  and L.  Davidovich,  Phys.  Rev.  A {\bf78},  022322
(2008).

\bibitem{22} J.  M.  Raimond,  M.  Brune,  and S.  Haroche,  Rev.  Mod.   Phys.  {\bf 73},
565 (2001).


\bibitem{Vidal}
G. Vidal and R. F. Werner, Phys. Rev. A {\bf 65}, 032314 (2002); M.
B. Plenio, Phys. Rev. Lett. {\bf 95}, 090503 (2005).

\bibitem{23} V. Coffman, J. Kundu, and W. K. Wootters, Phys. Rev. A {\bf 61}, 052306 (2000).

\end{thebibliography}
\end{document}